\documentclass[10pt,preprint,numberedappendix]{emulateapj_2004}
\usepackage{pdflscape}
\usepackage{graphicx}
\usepackage{color}
\usepackage{amsmath,amsthm,amssymb}
\usepackage{epsfig}
\usepackage{mathrsfs}
\usepackage{url}
\newcommand{\Msun}{\mbox{$M_{\odot}$}}
\newcommand{\Mjup}{\mbox{$M_{Jup}$}}
\newcommand{\gps}{\ensuremath{g_{\rm P1}}}
\newcommand{\rps}{\ensuremath{r_{\rm P1}}}
\newcommand{\ips}{\ensuremath{i_{\rm P1}}}
\newcommand{\zps}{\ensuremath{z_{\rm P1}}}
\newcommand{\yps}{\ensuremath{y_{\rm P1}}}

\begin{document}

\title{A Pan-STARRS + UKIDSS Search for Young, Wide Planetary-Mass Companions \\ 
in Upper Scorpius}
\author{Kimberly M. Aller \altaffilmark{1,2}, Adam L. Kraus\altaffilmark{1,3}, Michael C. Liu\altaffilmark{1}, William S. Burgett\altaffilmark{1}, Kenneth C. Chambers\altaffilmark{1}, Klaus W. Hodapp\altaffilmark{1}, Nick Kaiser\altaffilmark{1}, Eugene A. Magnier\altaffilmark{1}, Paul A. Price\altaffilmark{4}}

\altaffiltext{1}{University of Hawaii, Institute of Astronomy}
\altaffiltext{2}{Visiting Astronomer at the Infrared Telescope Facility, which
  is operated by the University of Hawaii under Cooperative Agreement
  no. NNX-08AE38A with the National Aeronautics and Space
  Administration, Science Mission Directorate, Planetary Astronomy
  Program.} 
\altaffiltext{3}{Harvard-Smithsonian Center for Astrophysics, 60
  Garden Street, Cambridge, MA 02138, USA}
\altaffiltext{4}{Department of Astrophysical Sciences, Princeton University, Princeton, NJ 08544, USA}

\submitted{ }
\journalinfo{\apj, Accepted June 17, 2013}


\begin{abstract}

We have combined optical and NIR photometry from Pan-STARRS~1 and
UKIDSS to search the young (5--10\,Myr) star-forming region of Upper
Scorpius for wide ($\approx$\,400--4000\,AU) substellar companions
down to $\sim$\,5\,\Mjup. Our search is $\approx$\,4\,mag deeper than
previous work based on 2MASS. We identified several candidates around
known stellar members using a combination of color selection and
spectral energy distribution fitting. Our followup spectroscopy has
identified two new companions as well as confirmed two companions
previously identified from photometry, with spectral types of M7.5-M9
and masses of $\sim$\,15--60\,\Mjup, indicating a frequency for such
wide substellar companions of $\sim$\,0.6\,$\pm$\,0.3\%. Both
USco\,1610$-$1913B and USco\,1612$-$1800B are more luminous than
expected for their spectral type compared with known members of Upper
Sco. HIP~77900B has an extreme mass ratio
($M_{2}/M_{1}$\,$\approx$\,0.005) and an extreme separation of
3200\,AU. USco\,1602$-$2401B also has a very large separation of
1000\,AU. We have also confirmed a low-mass stellar companion,
USco\,1610$-$2502B (730\,AU, M5.5). Our substellar companions appear
both non-coeval with their primary stars according to evolutionary
models and, as a group, are systematically more luminous than the
Upper Sco cluster sequence. One possible reason for these luminosity
discrepancies could be different formation processes or accretion
histories for these objects. 
\end{abstract}


\section{Introduction}

Recent advances in direct imaging techniques have led to the discovery
of planets in moderately wide ($\sim$\,10--100\,AU) orbits around
other stars, such as Formalhaut \citep{2008Sci...322.1345K}, HR~8799
\citep{2008Sci...322.1348M,2010Natur.468.1080M}, and $\beta$~Pic
\citep{2009AA...493L..21L}. Direct imaging surveys have also
discovered planetary-mass ($\lesssim$\,13\,\Mjup) companions with very
large ($\sim$\,200--500\,AU) orbital radii including
1RXS~J1609$-$2105B \citep[8\,\Mjup, 330\,AU;][]{2008ApJ...689L.153L},
CHXR~73B \citep[12\,\Mjup, 210\,AU;][]{2006ApJ...649..894L}, and
GSC~06214$-$00210B \citep[14\,\Mjup,
  330\,AU;][]{2011ApJ...726..113I}. The most extreme of such wide
companions is WD~0608-661B with a mass of 7\,\Mjup\ and a projected
separation of 2500\,AU \citep{2011ApJ...730L...9L}. It is difficult to
determine whether a planetary-mass companion at such a large distance
formed from a protoplanetary disk (and thus should be considered a
planet) or as a binary system (and should be called a brown
dwarf). Regardless of their origins, detailed spectroscopic and
photometric analysis of these directly imaged systems
\citep[e.g.][]{2008ApJ...689L.153L,2010ApJ...723..850B} can yield
insight into the properties (e.g. luminosity, temperature and mass) of
gas-giant planets, and thereby shed light on the over 500 radial
velocity and transiting exoplanets that lie within a few AU of their
host stars and therefore cannot be directly studied.

Determining the mass function, separation distribution, and frequency
of these wide planetary-mass companions will provide insight into
their formation. The precise boundary between planets and brown dwarfs
is still under debate. The most widely used definition adopts the
deuterium-burning limit of
$\approx$\,13\,\Mjup\ \citep[e.g.][]{2011ApJ...727...57S} to set the
boundary. Alternatively, the mass distribution of substellar
companions might provide a means to distinguish planets from brown
dwarfs, by shedding light on their formation process(es). The mass
function of companions to solar-type stars has two distinct
populations separated by a deficit of objects around
$\sim$\,30\,\Mjup\ (e.g. \citeauthor{2005prpl.conf.8252L},
\citeyear{2005prpl.conf.8252L}; a.k.a. the ``brown dwarf desert''),
suggesting that the population of objects with masses below this gap
might have a common origin and thus all be considered planets.

Similarily, the distribution of separations might provide valuable
clues. If wide substellar companions form like binary stars, we expect
that they may reside at separations as large as $\sim$\,1000~AU from
their primary star, as stellar binaries are observed at such large
separations. Such companions would represent the extreme low-mass end
of binary star formation \citep{2010ApJ...710.1375K}. Therefore we
would expect that their separation distribution would be similar to
that for stars and brown dwarfs \citep{2011ApJ...731....8K} such that
their distribution would be logarithmically flat (i.e. companion
masses are equally likely in $log$(separation)). If instead wide
substellar companions form like planets, we do not expect to find them
beyond a few hundred AU, because protoplanetary disks should not form
planets so far out.  \citet{2009ApJ...707...79D} find that even planet
formation in moderately wide orbits (35--100\,AU), such as in the case
of HR~8799, could not have formed via core accretion. Whether or not
the competing disk-instability model \citep{2001ApJ...563..367B} can
form planets at such large separations has not been well
explored. Disks have only been modeled to moderate radii
\citep[300\,AU; e.g.][]{2010ApJ...710.1375K,2010MNRAS.406.2279M}. In
addition, the typical sizes of circumstellar disks range from
$\sim$\,100--400\,AU \citep{2005AA...441..195V}, making in situ
formation by disk-instability at very wide separations, where there is
no disk at all, unlikely.

Substellar and planetary-mass companions are expected to cool and fade
quickly after their formation and therefore are most readily detected
in young ($\lesssim$\,10\,Myr) star-formation regions. We have used
the UKIDSS Galactic Cluster Survey (GCS) and the Pan-STARRS~1 (PS1)
3$\pi$\ Survey to search for wide planetary-mass companions in the
Upper Sco association. A similar search has previously been conducted
with 2MASS \citep{2007ApJ...662..413K}, but the 2MASS detection limit
($K$\,=\,14.3; $\sim$\,20\,\Mjup\ in Upper Sco) is $\sim$\,4\,mag
brighter than for UKIDSS ($K$\,=\,18.2; $\sim$\,5\,\Mjup\ in Upper
Sco). Combining both optical (PS1) and NIR (UKIDSS) data increases the
wavelength coverage, significantly improving the ability to reject
reddened background stars as potential planetary-mass companions. In
Section~\ref{sec:data} we discuss the UKIDSS GCS and the PS1
photometric data. In Section~\ref{sec:searchmethod} we describe our
search method and how we photometrically determine the spectral type
of our candidates. In Section~\ref{sec:results} we describe the
spectroscopic followup and our new substellar and a low-mass stellar
companions. Our dicussion is in Section~\ref{sec:discussion} and our
conclusions are in Section~\ref{sec:conclusion}.


\section{Survey Data} \label{sec:data}

\subsection{UKIDSS}
The United Kingdom Infrared Deep Sky Survey (UKIDSS) began in 2005 and
uses the 3.8\,m United Kingdom Infrared Telescope (UKIRT) located on
Mauna Kea \citep{2006AJ....131.1163S}. The UKIDSS project is defined
in \citet{2007MNRAS.379.1599L}. UKIDSS uses the UKIRT Wide Field
Camera \citep[WFCAM; ][]{2007AA...467..777C} and a photometric system
described in \citet{2006MNRAS.367..454H}. The pipeline processing and
science archive are described in \citet{2004SPIE.5493..411I} and
\citet{2008MNRAS.384..637H}. UKIDSS consists of five surveys: the
Galactic Clusters Survey (GCS), the Large Area Survey (LAS), the
Galactic Plane Survey (GPS), the Ultra-Deep Survey (UDS) and the Deep
Extragalactic Survey (DXS). The GCS covers $\approx$\,1400\,deg$^{2}$
of galactic star-formation regions and open clusters visible from the
Northern Hemisphere ($\delta$\,$\gtrsim$\,$-$30$^{\circ}$), including
the Upper Sco star-forming region, in 5 NIR bands, $ZYJHK$
($\approx$\,0.8\,$-$\,2.4\,$\mu$m) \citep{2007MNRAS.379.1599L}. This
survey is only $\sim$\,40\% complete (by area) and the 5\,$\sigma$
limiting magnitudes (Vega) in the observed area are $Z$\,=\,20.4,
$Y$\,=\,20.1, $J$\,=\,19.6, $H$\,=\,18.8, and $K$\,=\,18.2\,mag
\footnotemark\footnotetext{http://surveys.roe.ac.uk/wsa/dr9plus\_release.html}. In addition, we use data from the other UKIDSS
surveys (LAS, UDS, and DXS) to construct stellar spectral energy (SED)
templates of known dwarfs in order to estimate the spectral type of
our candidates. See Section~\ref{sec:sed_chi2},
Section~\ref{sec:sed_chi22}, and Appendix~\ref{appendix:sed_templates}
for more details.

For both our wide companion search and ultracool dwarf template
construction, we use the catalog data from the UKIDSS DR9
release. Also, all UKIDSS magnitudes are on the Vega system. We chose
good data as having the following properties: magnitude errors
$\le$\,0.2\,mag, not deblended and without saturated, almost-saturated
or bad pixels. These requirements remove most of the spurious
detections we may encounter near bright stars. Furthermore, we ignore
all detections $<$1\arcsec\ from the location of our primary star
because they likely correspond either to the actual primary star or
small artifacts within the star's point spread function (PSF). We note
that our completeness within 3\arcsec\ from the primary star is still
very low because any companion would be likely to be contaminated by
the relatively brighter primary star's PSF.

\subsection{Pan-STARRS~1}

Pan-STARRS~1 (PS1) is a 1.8\,m, wide-field telescope located on
Haleakal\={a} on the island of Maui, conducting a multi-wavelength,
multi-epoch, optical imaging survey \citep{2002SPIE.4836..154K}.  Its
large sky coverage ($\approx$\,30,000\,deg$^{2}$) coupled with its
\zps\ ($\lambda_{eff}$\,=\,866\,\AA) and \yps\ ($\lambda_{eff}$\,=\,962\,\AA)
filters provide both coverage of the same star-forming regions as
UKIDSS and greater sensitivity at longer wavelengths, where brown
dwarf and young gas giant planet spectra are brighter compared to the
shorter wavelengths. Our work is the first to use PS1 data to search
for young brown dwarfs and planetary-mass companions.

We use data from the PS1 3\,$\pi$\ survey, which began in 2010, both to
search for candidate companions and to construct spectral energy
distribution templates of ultracool dwarfs (see
Section~\ref{sec:sed_chi2}, Section~\ref{sec:sed_chi22}, and
Appendix~\ref{appendix:sed_templates}). The 3\,$\pi$\,Survey covers
$\approx$\,75$\%$ of the sky in 5 optical filters, \gps, \rps, \ips,
\zps, and \yps\ \citep{2012arXiv1203.0297T}. At each epoch a single
field is exposed for 43\,s in \gps, 40\,s in \rps, 45\,s in \ips,
30\,s in \zps, and 30\,s in \yps. The photometry from the reduced
multi-epoch data have been averaged to calculate mean magnitudes. The
predicted final limiting magnitudes, on the AB system, for each filter
are 23.4, 22.8, 22.2, 21.6, and 20.1\,mag in \gps, \rps, \ips, \zps,
and \yps, respectively
\citep{2009ApJ...704.1519D,2012arXiv1203.0297T}.

We chose good quality data according to the photometric quality flags
set in the PS1 Desktop Virtual Observatory (DVO) database
\citep{2006amos.confE..50M}. We queried the DVO catalogs for
3\,$\pi$\ survey and selected objects with the following attributes:
fits a PSF model (is not extended); is not saturated; has a good sky
measurement; is not likely a cosmic ray, a diffraction spike, a ghost
or a glint; does not lie between the image chips; and has the quality
flag {\tt psf\_qf}\,$\ge$\,0.9 to ensure that at least 90\% of the
object is unmasked. Furthermore, we require objects to be detected at
least twice in a single night in at least one of the five filters to
remove potential spurious sources that would only appear as single
detections. Finally, we require that a single bandpass measurement
error be $\le$\,0.2\,mag in order to use that bandpass. We note that
we obtained photometry from the PS1 database prior to the updated
photometric calibrations \citep{2012ApJ...756..158S}. Because the
previous database had misreported some photometric errors as below
0.01\,mag, we have capped the reported photometric error at
0.01\,mag. All PS1 photometry tabulated is from the previous database,
for consistency with our actual search and analysis.


\section{Candidate Selection} \label{sec:searchmethod}
We combined the UKIDSS GCS and PS1 3\,$\pi$\ catalogs to search for
objects located within 1$-$30\arcsec\ of known Upper Sco members using
TOPCAT\footnotemark
\footnotetext{\url{http://www.starlink.ac.uk/topcat}}
\citep{2005ASPC..347...29T}. Bona fide members are taken from
\citet{1989AA...216...44D}, \citet{1999PhDT........89K},
\citet{1994AJ....107..692W}, \citet{1999AJ....117..354D},
\citet{2000AJ....120..479A}, \citet{2002AJ....124..404P},
\citet{2004AJ....127..449M}, \citet{2006MNRAS.373...95L},
\citet{2008ApJ...688..377S} and \citet{2011MNRAS.416.3108R}.
\citet{2007ApJ...662..413K} compiles these lists for all members
determined prior to the paper. Our final input list includes 673
spectroscopically confirmed Upper Sco members.

The decreased detection efficiency within 3\arcsec\ limits our search
to companions with projected separations greater than
$\sim$\,400\,AU. Furthermore, the UKIDSS $K$-band sensitivity can
detect companions down to $\sim$\,5\,\Mjup, assuming an age of
5--10\,Myr for Upper Sco \citep{2000ApJ...542..464C}.

\subsection{Color Selection}

\begin{figure}
  \begin{center}
    \includegraphics[angle=90,width=.8\textwidth]{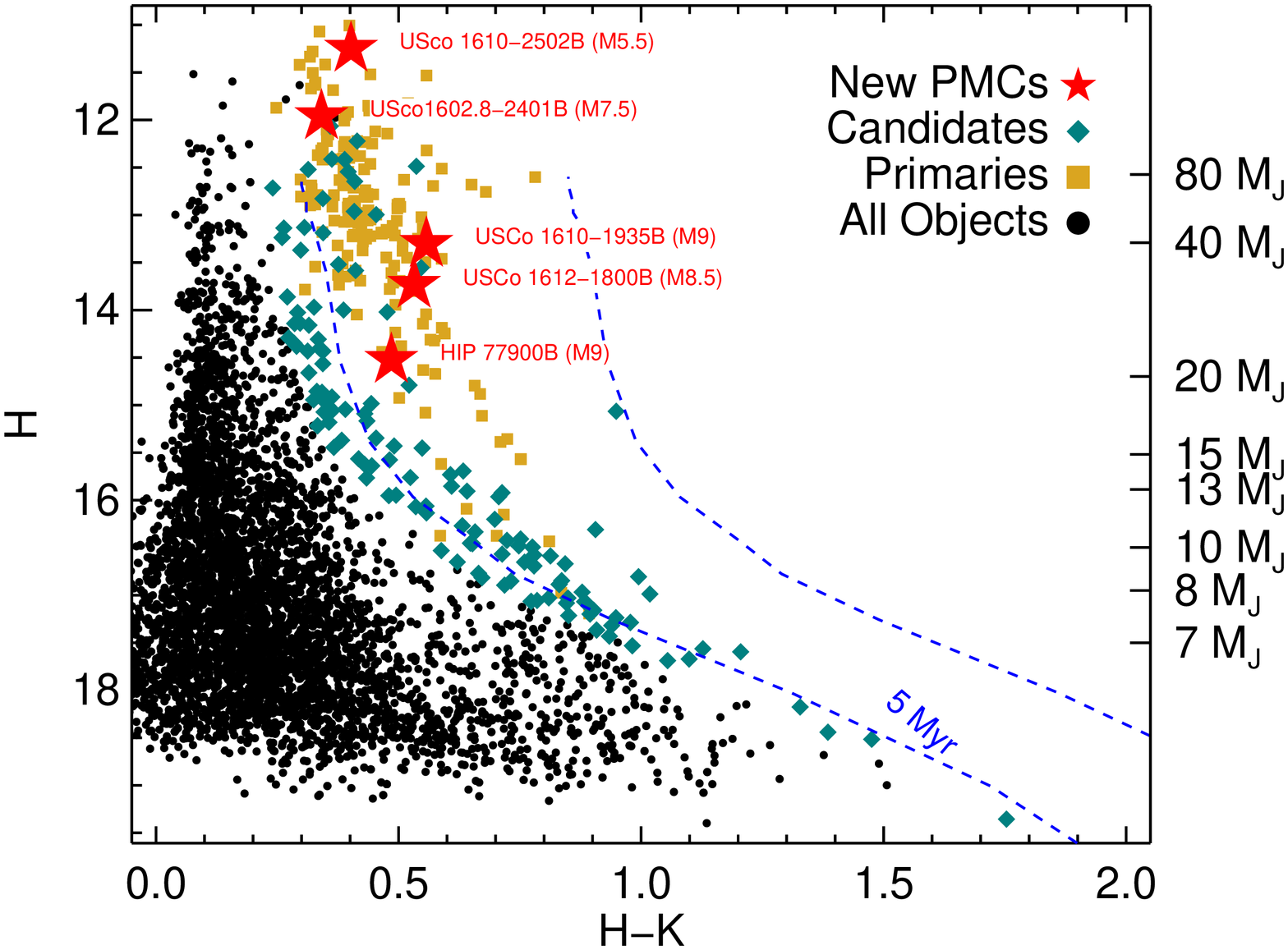}
    \caption{A color-magnitude diagram using the UKIDSS $H$ and $K$
      magnitudes for all point sources within $\le$~30\arcsec\ of a
      known Upper Sco member. Known Upper Sco members are shown with
      orange squares and stand out from the background stars.  The
      mass scale is shown using the Lyons/DUSTY evolutionary models
      \citep{2000ApJ...542..464C} for a 5~Myr sequence (\emph{right
        y-axis}) where the $H-K$ colors are given in the models. Our
      initial candidates (prior to the SED fit) are the teal diamonds
      which roughly follow the model sequence. The dashed \emph{blue}
      lines outline the DUSTY 5~Myr sequence for Upper Sco where the
      two lines roughly encompass the observed spread in the sequence
      from the primary stars. The four new substellar companions
      (M7.5--M9, $\sim$\,15--60\,\Mjup) and one new low-mass stellar
      companion (M5.5, $\sim$\,0.1\,\Msun) are the \emph{red
        stars}. \label{fig:f1}}
  \end{center}
\end{figure}

Our initial candidate selection used NIR colors to isolate candidates
that lie along the Upper Sco color-magnitude sequence. This
significantly reduces the obvious background objects with neutral
colors. We then selected substellar candidates using UKIDSS $H$ and
$K$ photometry, which are nearly complete for Upper Sco. Our
candidates were selected to lie above a diagonal line which roughly
traces the cluster sequence and to be fainter than $H$\,=\,12
\citep[$\sim$\,90\,\Mjup;][]{2000ApJ...542..464C}. All companions
brighter than this limit should have been detected by
\citet{2009ApJ...703.1511K}. We empirically defined this line by
horizontally shifting the evolutionary model tracks until they
bracketed the edge of the observed primary star sequence
(Figure~\ref{fig:f1}).

\subsection{Spectral Types from SED Fitting} \label{sec:sed_chi2}
These initial color-selected candidates were then fit using our SED
template library to estimate their spectral type. See
Appendix~\ref{appendix:sed_templates} for a description of our SED
templates. We performed a $\chi^2$ minimization to determine the
spectral type using the available PS1 and UKIDSS photometry for each
of our candidates. Our $\chi^2$ minimization took into account
uncertainties in both the candidate data and the templates using the
following weight, $w_{i}$ for each filter, $i$:
\begin{equation}
  w_{i} = (\sigma_{obs,i}^{2} + \sigma_{SED,i}^{2})^{-1}
\end{equation}
where $\sigma_{obs,i}$ is the magnitude error in the candidate data
and $\sigma_{SED,i}$ is the magnitude error in the SED template. We
set the magnitude error in the SED templates to a constant value of
0.1\,mag in order to prevent the SED template with the largest
uncertainties (i.e. those constructed by averaging open cluster or
Upper Sco members) from returning the minimum $\chi^{2}$. This was an
issue because the individual photometric uncertainties are
$\sim$\,0.01\,mag, whereas the photometric scatter within each
spectral type in either the open clusters or Upper Sco is around
0.5\,mag. Then we calculated the distance modulus of each candidate
relative to {\it each} SED template, $DM_{j}$, by minimizing the
$\chi^2$ for the distance:
\begin{equation} \label{eqn:dm}
  DM_{j} = \dfrac{\displaystyle \sum\limits_{i=0}^{n} w_i (m_{obs,i} -
  m_{SED,ij})}{\displaystyle \sum\limits_{i=0}^{n}w_i}
\end{equation}
where $j$ is the SED template, $i$ is the filter, $m_{obs,i}$ is the
observed magnitude in a filter, $m_{SED,i}$ is the magnitude of the
SED template, $n$ is the total number of filters, and $w_i$ is the
weight, from the previous equation.

Since the SED templates can have a different number of filters, we
determined the reduced $\chi^{2}$ between the candidate data and
templates in order to compare the goodness of fit between different
templates. We also required that each template have UKIDSS photometry
because our initial color selection of candidates is done in the
NIR. Therefore, our final $\chi^{2}$ fitting matched the candidate
data with all of the SED templates with any NIR photometry and
measurements in at least three filters in common. We then determined
the relative distance modulus to each of these SED templates
(Equation~\ref{eqn:dm}) and the associated reduced $\chi^{2}$.

Finally, the best fit SED was chosen based on the minimum reduced
$\chi^2$. The absolute magnitude to spectral type relations (see
Appendix~\ref{appendix:sed_templates}) convert the relative distance
modulus determined in our $\chi^{2}$ fitting to an absolute scale.

\subsection{Spectral Type Uncertainties} \label{sec:sed_chi22}
There is an uncertainty in the spectral type estimates from the SED
fitting due to both the intrinsic scatter in the SEDs of field objects
of a given spectral type and the measurement uncertainties in both the
candidate data and the SED templates. To incorporate this uncertainty
in our final distance estimate, we carried out a Monte Carlo
simulation to find the distribution of best-fit templates. For each
candidate, we perturbed the photometry in each filter by drawing from
a Gaussian distribution corresponding to the magnitude errors. We
varied the SED template photometry in the same manner. For each
realization (out of a total of 5000), we recomputed the best-fit
distance and spectral type, resulting in a distribution of best fit
spectral type and distance. We generated 5000 iterations, because the
median distance and spectral type from the ensemble of fits converged
for a tested set of objects with a variety of measurement errors in
\yps.

The final best fit spectral type is the median from the Monte Carlo
simulation and the best fit template is the most probable template
(according to the Monte Carlo distribution) with that spectral
type. Note that the SED fitting routine may select different SEDs with
the same spectral type. Therefore, a spectral type uncertainty of zero
only means that the object had best fitting SED templates with the
same spectral type, not necessarily a single SED template. In these
cases we adopted a spectral type uncertainty according to the
available templates, usually approximately one spectral
subclass. Furthermore, the spectral type estimates are quantized and
thus we only have a total of 34 possible spectral types. For our
purposes, the spectral type uncertainty simply serves to further
distinguish the quality of the template fit and an object's candidacy.

\subsection{Final Candidate Selection}
We selected the final candidates combining the reduced $\chi^{2}$
($\chi$$^{2}_{\nu}$\,$<$\,5) and spectral type cut ($>$\,M7) from the
SED fitting, a visual check on the UKIDSS and/or PS1 images, and their
observability with IRTF SpeX according to their magnitude
($J$\,$\lesssim$\,18). Although the SED template fitting routine
efficiently removes many of the reddened background stars that pass
the initial NIR color selection, some still may pass. Our criteria
will miss objects without at least three-band photometry, but it gives
us a relatively pure candidate sample for followup spectroscopy in
comparison to a solely color-selected one. We started with a total of
285 color-selected candidates. There were a total of 30 candidates
remaining after the SED fitting for spectral type and our $\chi^{2}$
cut.


\section{Results} \label{sec:results}

\subsection{Spectroscopic Followup}

In principle proper motions could confirm candidates as
comoving. However the predicted uncertainty in the UKIDSS proper
motions is only $\sim$\,13\,mas\,yr$^{-1}$. This precision is
insufficient for Upper Sco because of its very low proper motion
\citep[$\mu_{\alpha}$,$\mu_{\delta}$ = $-$10,$-$25\,mas\,yr$^{-1}$;
][]{1999AJ....117..354D}. Therefore, we require spectroscopy to
confirm our candidates as true companions.

We obtained spectroscopic followup using SpeX
\citep{2003PASP..115..362R}, a medium-resolution, near-IR spectrograph
(0.8--2.5\,$\mu$m) on the 3-meter NASA Infrared Telescope Fascility
(IRTF) on Mauna Kea. The low resolution (LowRes15) prism mode with a
0.8\arcsec\ slit (R\,$\equiv$\,$\lambda/\Delta\lambda$\,$\approx$\,100) is
well-suited to spectroscopic confirmation of our substellar companions
which are expected to have spectral types of late-M to mid-L. Even at
this low resolution, the spectra of substellar companions will have broad
water band and molecular features that distinguish them from background
stars. We can also easily distinguish a triangular $H$-band continuum,
a feature which is characteristic of young low-mass objects
\citep{2001MNRAS.326..695L,2007ApJ...657..511A}.

We observed a total of 6 candidates on 2011 June 19-22 (UT) in cloudy
conditions with average seeing ($\sim$\,0.8\arcsec). On 2012 July 5 we
observed one candidate in poor seeing ($\sim$\,1.5\arcsec). On July
7-8 we observed another 7 candidates in excellent seeing
($\sim$\,0.5\arcsec). On 2013 April 16 we reobserved two companions to
obtain higher signal-to-noise spectra. Each observation used the
standard ABBA nod pattern for sky subtraction. We observed an A0V
standard star following each candidate (with the exception of a few
shared standards for nearby targets) then took wavelength and
flatfield calibrations immediately afterward. We reduced the data
using version 3.4 of the SpeXtool package \citep{2003PASP..115..389V,
  2004PASP..116..362C}. Table~\ref{table:tab1} tabulates the
observation details.

\subsection{Spectral Type}

The spectral resolution (R\,$\approx$\,100) is too low to determine
the spectral type using gravity/age independent flux indices
\citep{2007ApJ...657..511A}. Instead we first fit each spectrum to the
a collection of ultracool dwarf spectra in the IRTF/SpeX SpeX Prism
Library \footnotemark
\footnotetext{\url{http://pono.ucsd.edu/\textasciitilde
    adam/browndwarfs/spexprism}}. The adopted spectral type for our
candidates is that of the best fitting object. We adopt a spectral type
uncertainty of half a spectral subclass (e.g. M9\,$\pm$\,0.5) which
encompasses both the uncertainty in the spectral type of the SpeX
Prism Library objects and cases where the candidate may fit to more
than one object.

\begin{figure}
  \begin{center}
    \includegraphics[width=0.7\textwidth]{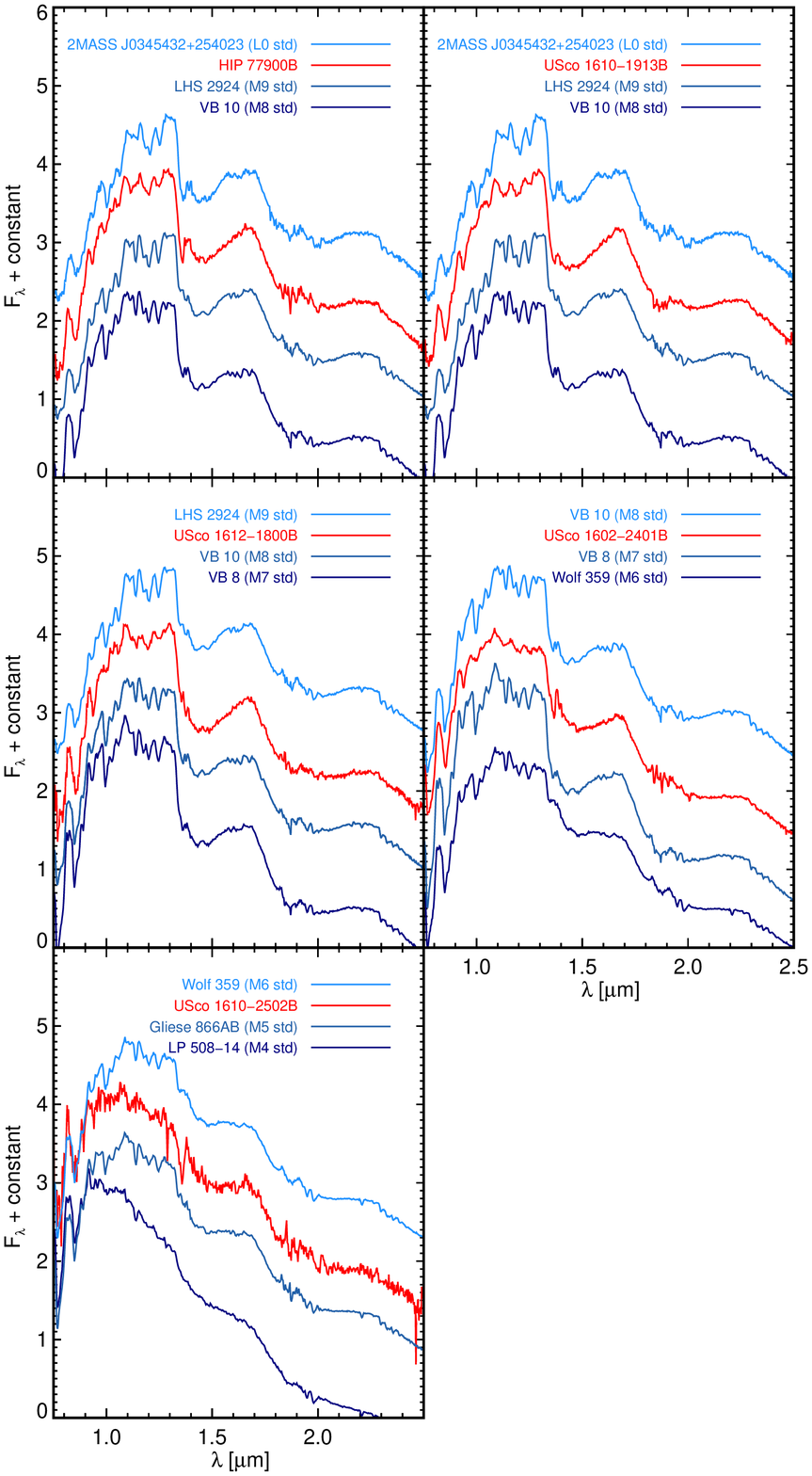}
    \caption{Comparison of our new companions to optical spectral
      standards for M~dwarfs \citep{1991ApJS...77..417K} and L~dwarfs
      \citep{1999ApJ...519..802K}: LP~580-14
      \citep[M4,~][]{2004AJ....127.2856B}, Gliese~866AB
      \citep[M5,~][]{2008ApJ...681..579B}, Wolf~359
      \citep[M6,~][]{2008ApJ...681..579B}, VB~8
      \citep[M7,~][]{2008ApJ...681..579B}, VB~10
      \citep[M8,~][]{2004AJ....127.2856B}, LHS~2924
      \citep[M9,~][]{2006AJ....131.1007B}, and 2MASS~J0345432+254023
      \citep[L0,~][]{2006AJ....131.1007B}. In the standards, the slope
      of both the $H$ and $K$ band continuum steadily changes from
      negative to positive with increasing spectral type. We
      determined the spectral type by visually comparing our
      companions with these standards according to the $H$ and $K$
      continuum shape. We note that the spectra of the spectral
      standards and our young companions are not necessarily the same
      because of their age difference. The $H$-band continuum is
      notably triangular in young low-mass objects whose lower gravity
      leads to decreased H$_{2}$ collision induced absorption making
      the H$_{2}$O absorption more prominent
      \citep{2001MNRAS.326..695L,2007ApJ...657..511A,2006ApJ...639.1120K}. Our
      companions are plotted in the following order: HIP\,77900B,
      USco\,1610$-$1913B, USco\,1612$-$1800B, USco\,1602$-$2401B,
      USco\,1610$-$2502B. \label{fig:f2}}
  \end{center}
\end{figure}

In addition, we compared our candidates to optical M and L~dwarf
standards observed with IRTF/SpeX also in prism mode, a common
method to determine spectral type in the NIR
\citep[e.g.][]{2003ApJ...593.1093L,2007AJ....134..411M}. These M~dwarf
standards are from the spectral classification scheme of
\citet{1991ApJS...77..417K} and the L~dwarf standards of
\citet{1999ApJ...519..802K}. The M--L~dwarf NIR spectral sequence in
low resolution (R\,$\approx$\,100) is characterized by water absorption
at 1.4\,$\mu$m and 1.8\,$\mu$m. Thus the $H$-band and $K$-band slopes
change with spectral type. Comparisons with the M--L~dwarf optical
standards yield the same spectral type as comparisons with both the
SpeX Prism Library and young M~dwarfs, although there are
discrepancies due to the youth of our companions. Figure~\ref{fig:f2}
compares each of our companions with the M and L~dwarf standards.

Finally, we compared our candidates to young M~dwarfs from
\citet{2007AJ....134..411M} with spectral types straddling our
candidates' best fit spectral type from the Prism Library
comparison. For each candidate, the best fitting young M~dwarf is
clearly the best match compared to the other young M~dwarfs with a
spectral type difference of just half a subclass. Therefore our
adopted spectral type uncertainty of half a subclass is consistent.
The final spectral type is that of the best matching young
M~dwarf. Our new companions have spectral types of M9, M9, M8.5, M7.5,
and M5.5. We tabulate their properties in Table~\ref{table:tab2}.

\subsection{Companion Physical Properties} \label{sec:physicalproperties}

We calculated both the effective temperatures ($T_{eff}$) and the
bolometric magnitudes ($M_{bol}$) of our companion discoveries and
their primaries using empirical relationships from the literature. We
adopted a spectral type uncertainty of half a subclass for all primary
stars, except in the case of HIP\,77900 where we assumed an
uncertainty of one subclass. For all objects we performed a Monte
Carlo simulation to derive the physical properties and their 68$^{th}$
percentile confidence limits. For each object, we perturbed its
spectral type and the resulting $T_{eff}$ by drawing from a Gaussian
distribution corresponding to the uncertainty in each parameter. The
$T_{eff}$ has uncertainties due to both the spectral type
uncertainties and the conversion from spectral type to $T_{eff}$. Our
methods to calculate $T_{eff}$, $M_{bol}$, and mass are slightly
different depending on the spectral type of the object.

For objects with spectral type $\ge$\,M5, we determined $T_{eff}$ from
the observed spectral type using the young M~dwarf scale of
\citet{2003ApJ...593.1093L}. We note that the
\citet{2003ApJ...593.1093L} scale is tailored for young stars and
brown dwarfs, like our companions. For comparison, we also converted
spectral type to $T_{eff}$ using the \citet{2004AJ....127.3516G}
relationship for old field dwarfs with spectral types later than M6
(with an uncertainty of 124\,K). We quote the $T_{eff}$ values from
both methods in Table~\ref{table:tab2} and adopt an intrinsic
$T_{eff}$ uncertainty of 124\,K for both conversions. The $T_{eff}$
uncertainties in the table also take into account the uncertainty in
spectral types. We used both $T_{eff}$ values to determine the mass
using the \citet{2000ApJ...542..464C} Lyon/DUSTY evolutionary models
for both 5\,Myr and 10\,Myr, corresponding to the age range of Upper
Sco.  To compute bolometric luminosity, we used bolometric corrections
as a function of spectral type from \citet{2004AJ....127.3516G} to
convert from $K$-band magnitude to $M_{bol}$.

For stars with earlier spectral types (M0--M5), we used the empirical
relationship between temperature and spectral type from
\citet{2003ApJ...593.1093L} and then derived the mass using $T_{eff}$
from the \citet{1998AA...337..403B} Lyon/NextGen evolutionary
models. Then we used the spectral type to select the $V$-band
bolometric corrections from \citet{1982_Schmidt-Kaler} and the
photospheric $V-K$ color from \citet{1988PASP..100.1134B}. We
converted from the $V-K$ color to 2MASS $K_{S}$ using the color
transformation in \citet{2001AJ....121.2851C}. We then used these
quantities with the observed $K_{S}$ magnitude to calculate $M_{bol}$.

For the earliest type stars we used the calibrations between spectral
type and $T_{eff}$ from \citet{1982_Schmidt-Kaler} for spectral types
of B8--K7 and from \citet{1995ApJS..101..117K} for spectral types
$<$\,B8. We then derived the mass using $T_{eff}$ from the
\citet{2000AA...358..593S} evolutionary models for both 5 and
10\,Myr. We assumed an $T_{eff}$ uncertainty of 100\,K from the
spectral type to $T_{eff}$ conversion. We used the observed spectral
type to select $V$-band bolometric corrections using relationships
from \citet{1982_Schmidt-Kaler} for spectral types of B8--K7 and
\citet{1995ApJS..101..117K} for spectral types $<$\,B8. To select
$V-K$ color we used the relationships between spectral type from
\citet{1988PASP..100.1134B}. We then converted from $V-K$ to 2MASS
$K_{S}$ using the color transformation from
\citet{2001AJ....121.2851C}. The bolometric luminosity was then
calculated using the observed $K_{S}$ magnitude.

We also calculated the projected separation for our companions
assuming the mean distance to Upper Sco, 145\,$\pm$\,2\,pc
\citep{1999AJ....117..354D}, and its depth on the sky, resulting in a
final distance of 145\,$\pm$\,15\,pc. Furthermore, we ignored
extinction in Upper Sco because the extinction is small for all of our
objects ($A_{V}$\,$\lesssim$\,1), and thus should not significantly
affect the companion properties derived from the NIR spectra.

Finally, we searched for signatures of disks using {\it WISE}
photometry \citep{2010AJ....140.1868W} of the primary
stars. \citet{2012ApJ...758...31L} suggests that 2MASS ($K_{S}$) and
{\it WISE} ($W3$ and/or $W4$) photometry can differentiate stars with
disks from those without disks. Stars of spectral type earlier than M4
with $K_{S}-W3$\,$\gtrsim$\,1 and $K_{S}-W4$\,$\gtrsim$\,1 should have
a disk. They may also show a wider range of luminosities, due to
thermal emission from the disk contaminating the $K_{S}$ magnitudes
used to compute $M_{bol}$. We found evidence for a debris/evolved
transitional disk around the USco\,1612$-$1800 system and marginal
evidence for a debris/evolved transitional disk around
USco\,1610$-$1913. The other three systems showed no evidence for a
disk.

Table~\ref{table:tab2} tabulates the photometry, the spectral type and
the final derived physical properties for the five new companions and
their primaries. Figure~\ref{fig:f8} shows the HR diagram for all
previously known members of Upper Sco and our new companions. See
Section~\ref{sec:discussion} for discussion.

\subsection{New Companions}

\begin{figure}
  \begin{center}
    \includegraphics[height=0.4\textwidth]{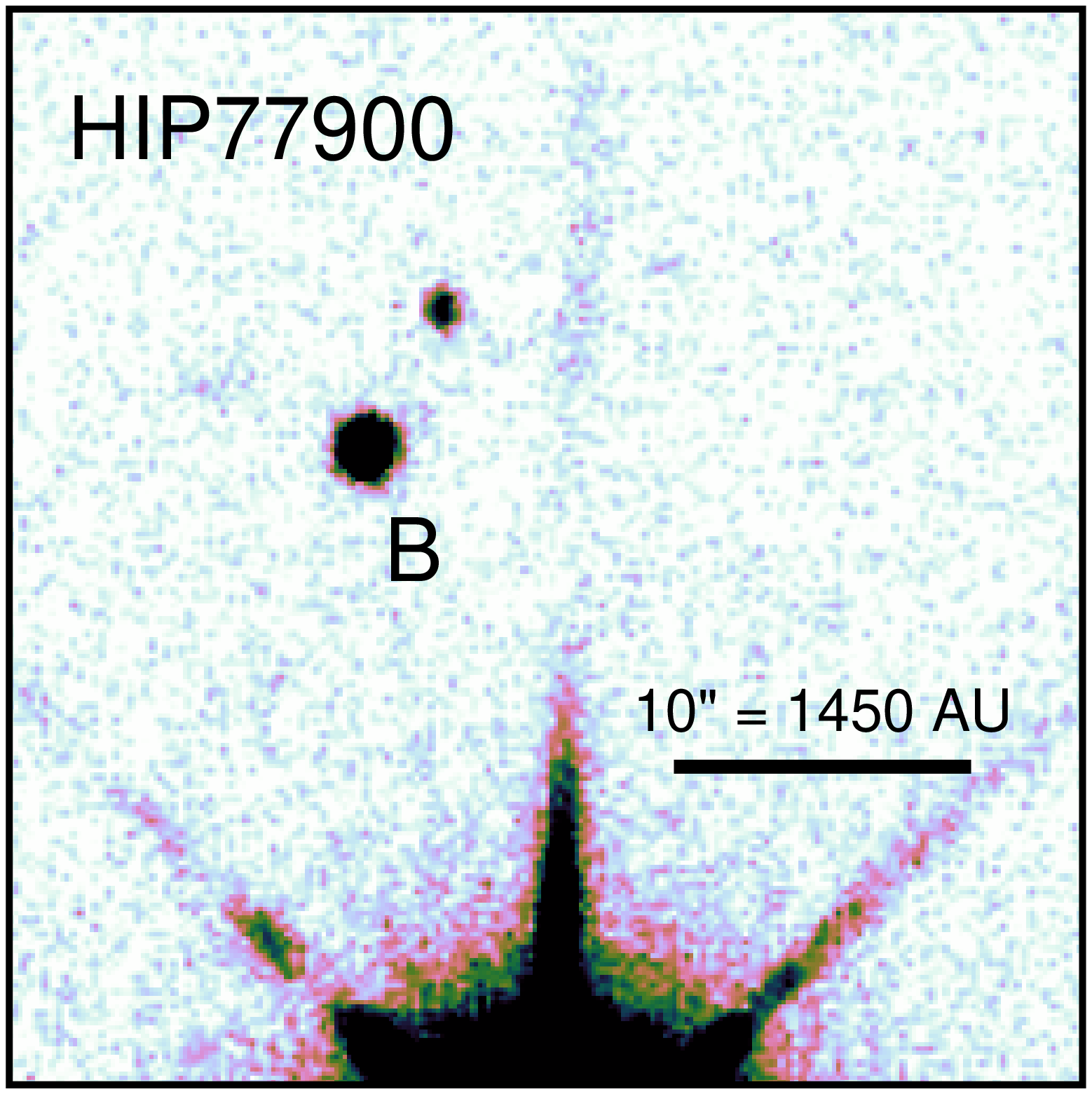}
    \includegraphics[height=0.4\textwidth]{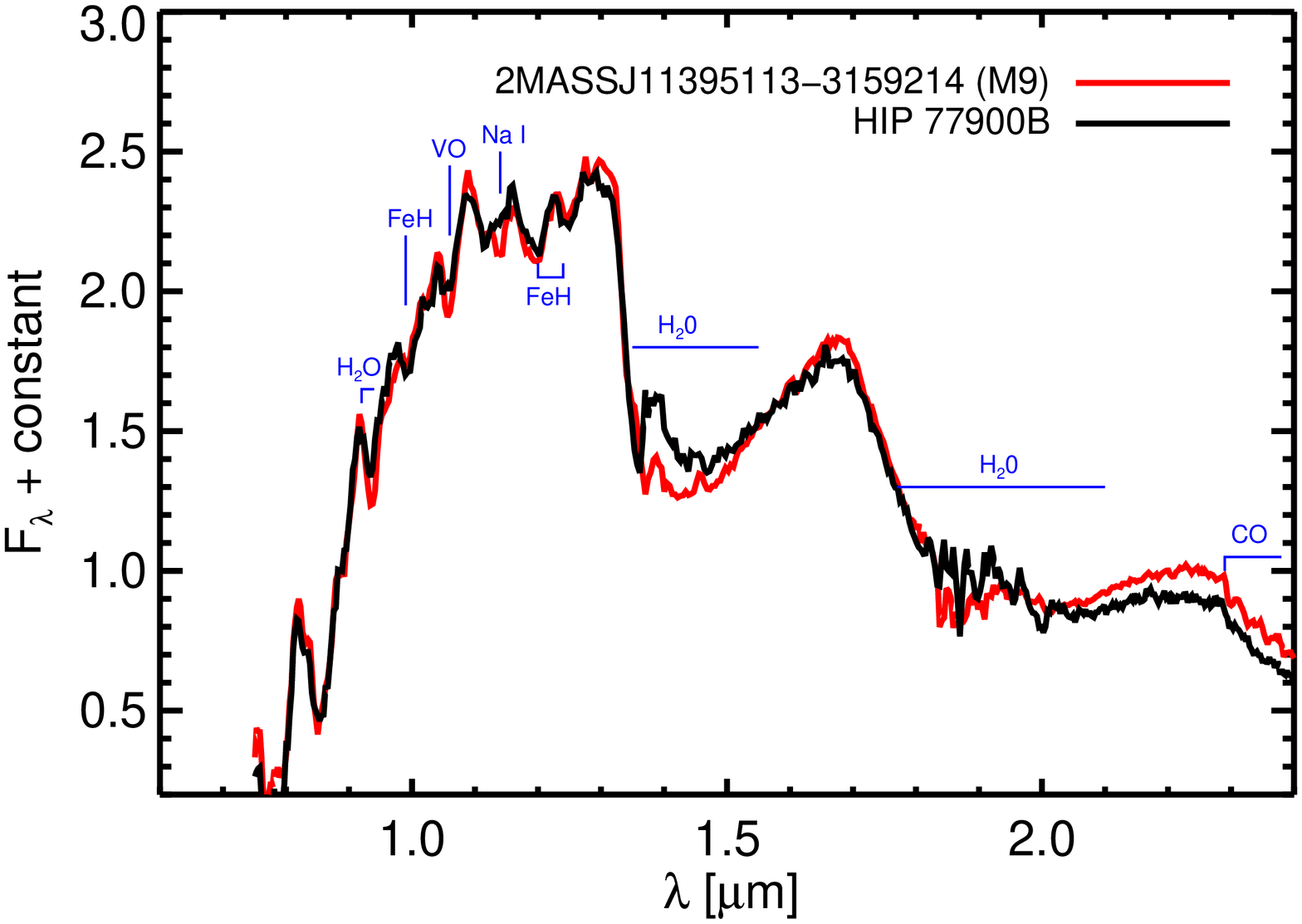}
    \caption{HIP\,77900B has a projected separation of
      $\approx$\,3100\,AU and a spectral type of M9
      ($\approx$\,19\,\Mjup). \emph{LEFT} -- The finder chart (UKIDSS
      $K$ and width of 30\arcsec) with the letter B identifying the
      companion. North is \emph{up} and East is
      \emph{left}. \emph{RIGHT} -- IRTF SpeX spectrum of HIP\,77900B
      compared to the young (8--12\,Myr) M9 in TW Hyrae
      \citep[2MASS~J11395113$-$3159214;]{2007ApJ...669L..97L}. \label{fig:f3}}
  \end{center}
\end{figure}

\begin{figure}
  \begin{center}
    \includegraphics[height=0.4\textwidth]{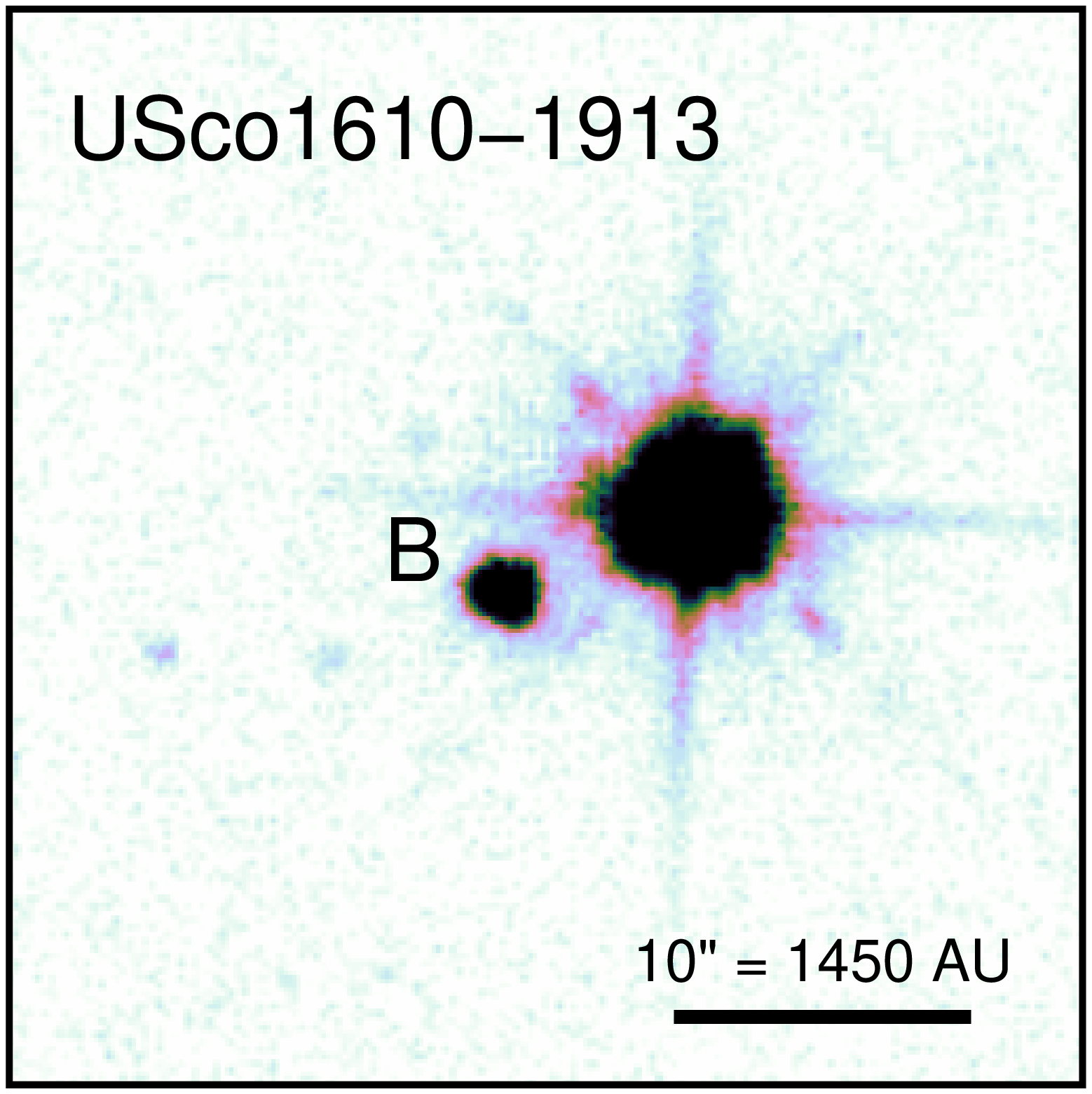}
    \includegraphics[height=0.4\textwidth]{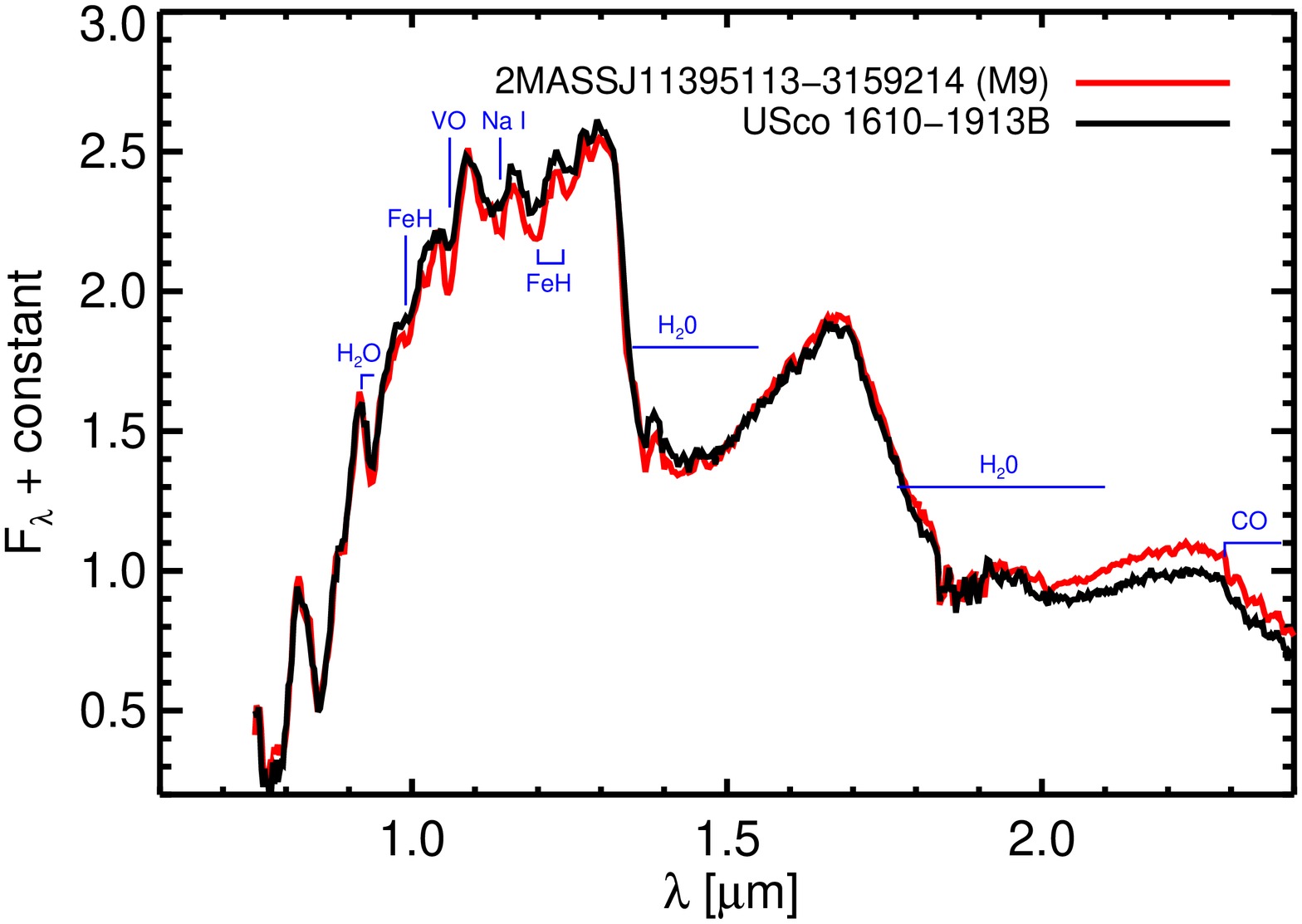}
    \caption{USco\,1610$-$1913B has a projected separation of
      $\approx$800\,AU and a spectral type of M9
      ($\approx$19\,\Mjup). \emph{LEFT} -- The finder chart (UKIDSS
      $K$ and width of 30\arcsec) with the letter B identifying the
      companion. North is \emph{up} and East is
      \emph{left}. \emph{RIGHT} -- IRTF SpeX spectrum of
      USco\,1610$-$1913B compared to the young (8--12\,Myr) M9 in TW
      Hydrae
      \citep[2MASS~J11395113$-$3159214;]{2007ApJ...669L..97L}. \label{fig:f4}}
  \end{center}
\end{figure}

\begin{figure}
  \begin{center}
    \includegraphics[height=0.4\textwidth]{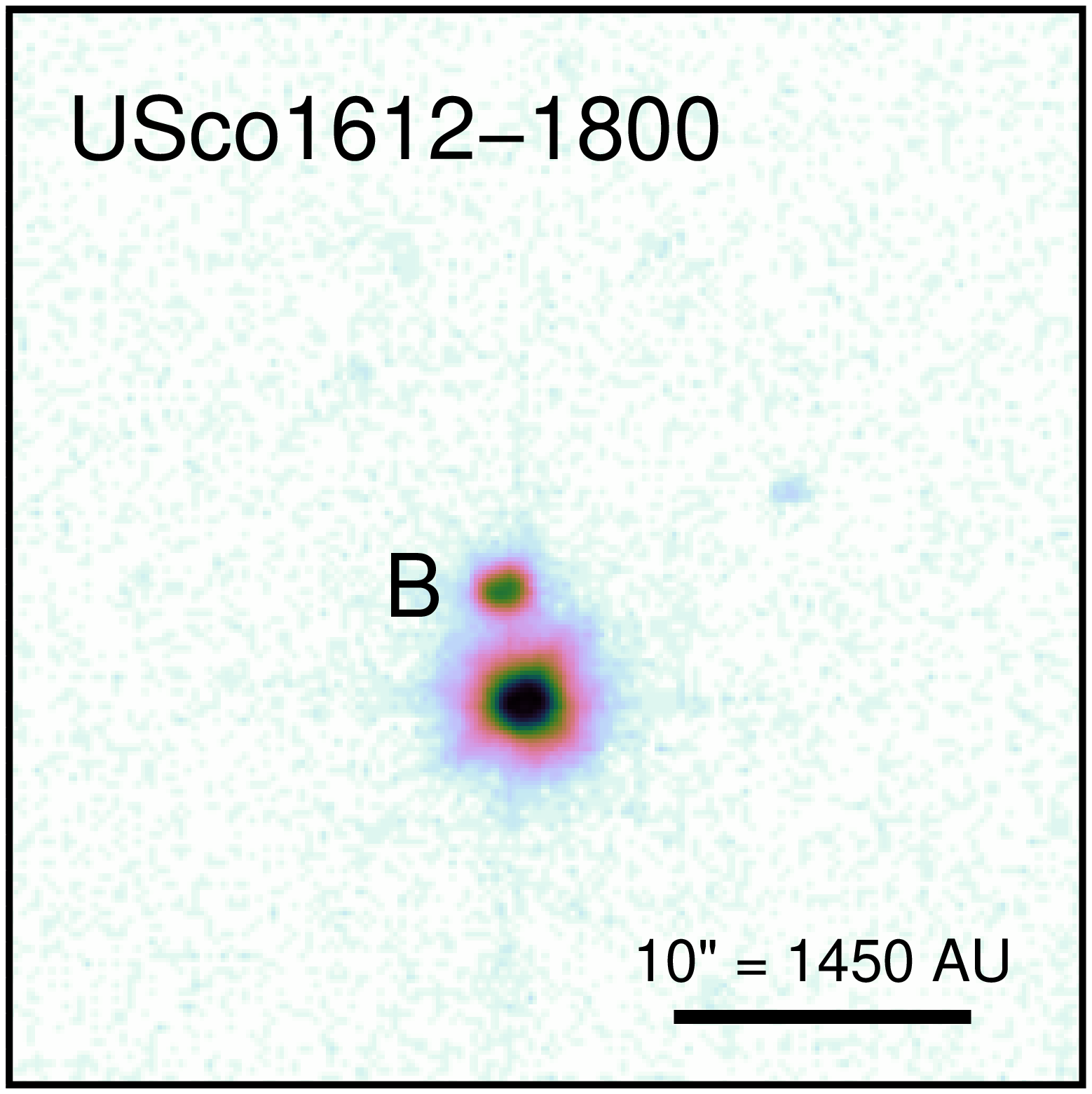}
    \includegraphics[height=0.4\textwidth]{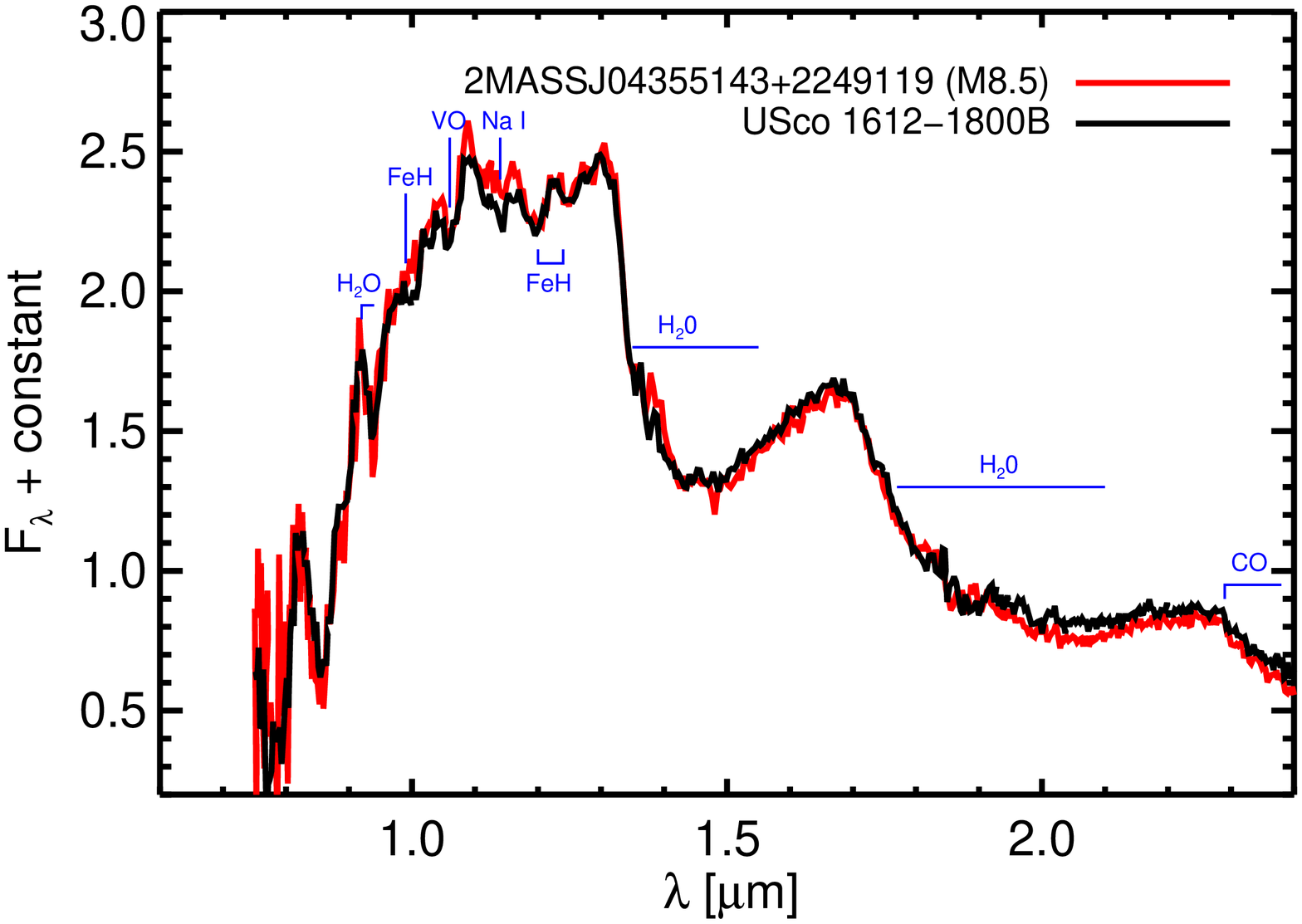}
    \caption{USco\,1612$-$1800B has a projected separation of
      $\approx$\,400\,AU and a spectral type of M8.5
      ($\approx$\,23\,\Mjup). \emph{LEFT} -- The finder chart (UKIDSS
      $K$ and width of 30\arcsec) with the letter B identifying the
      companion. North is \emph{up} and East is
      \emph{left}. \emph{RIGHT} -- IRTF SpeX spectrum of
      USco\,1612$-$1800B compared to the young (1--2\,Myr) M8.5 in
      Taurus \citep[2MASS~J04355143+2249119;
      ][]{2007AJ....134..411M}. \label{fig:f5}}
  \end{center}
\end{figure}

\begin{figure}
  \begin{center}
    \includegraphics[height=0.4\textwidth]{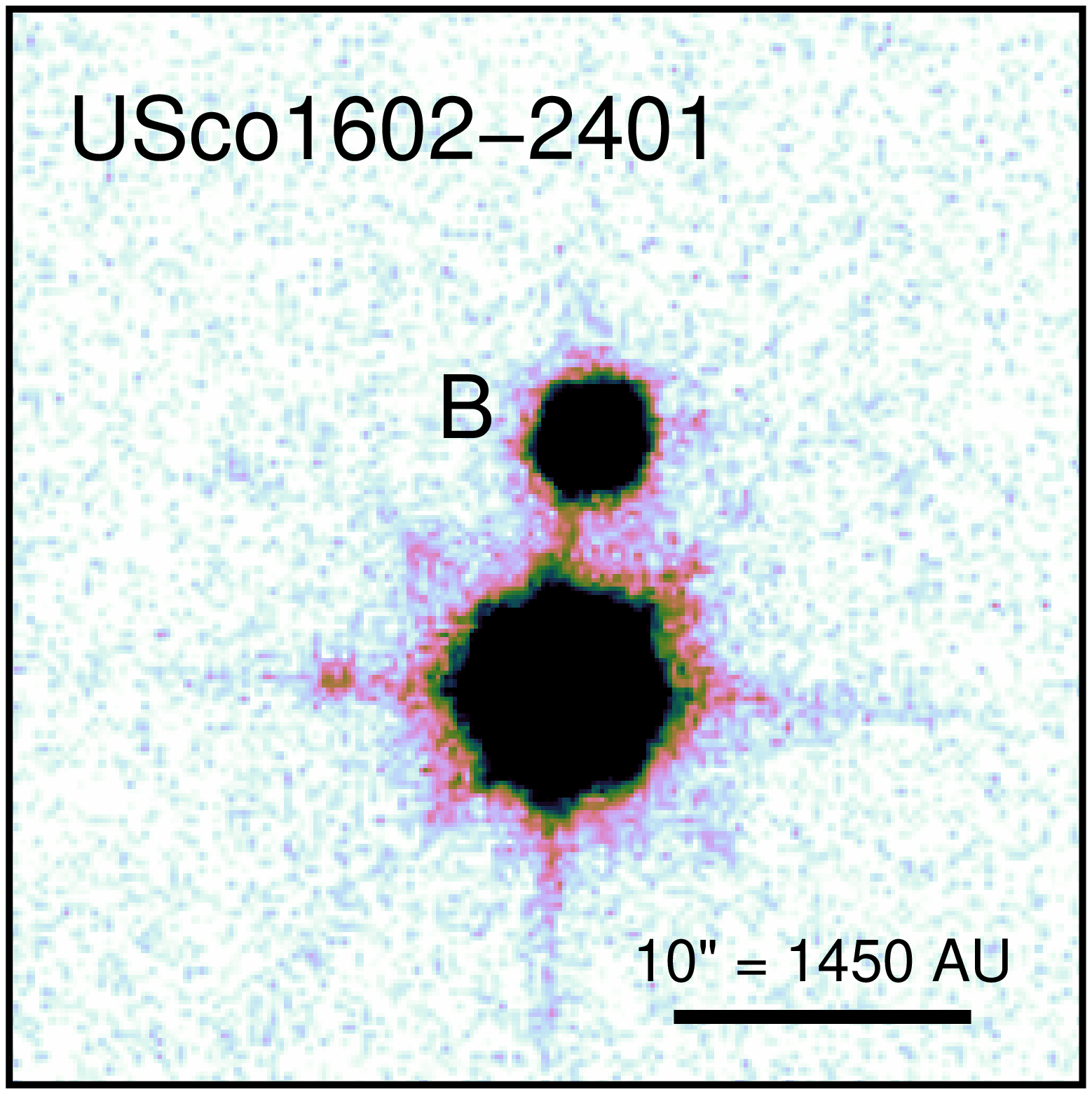}
    \includegraphics[height=0.4\textwidth]{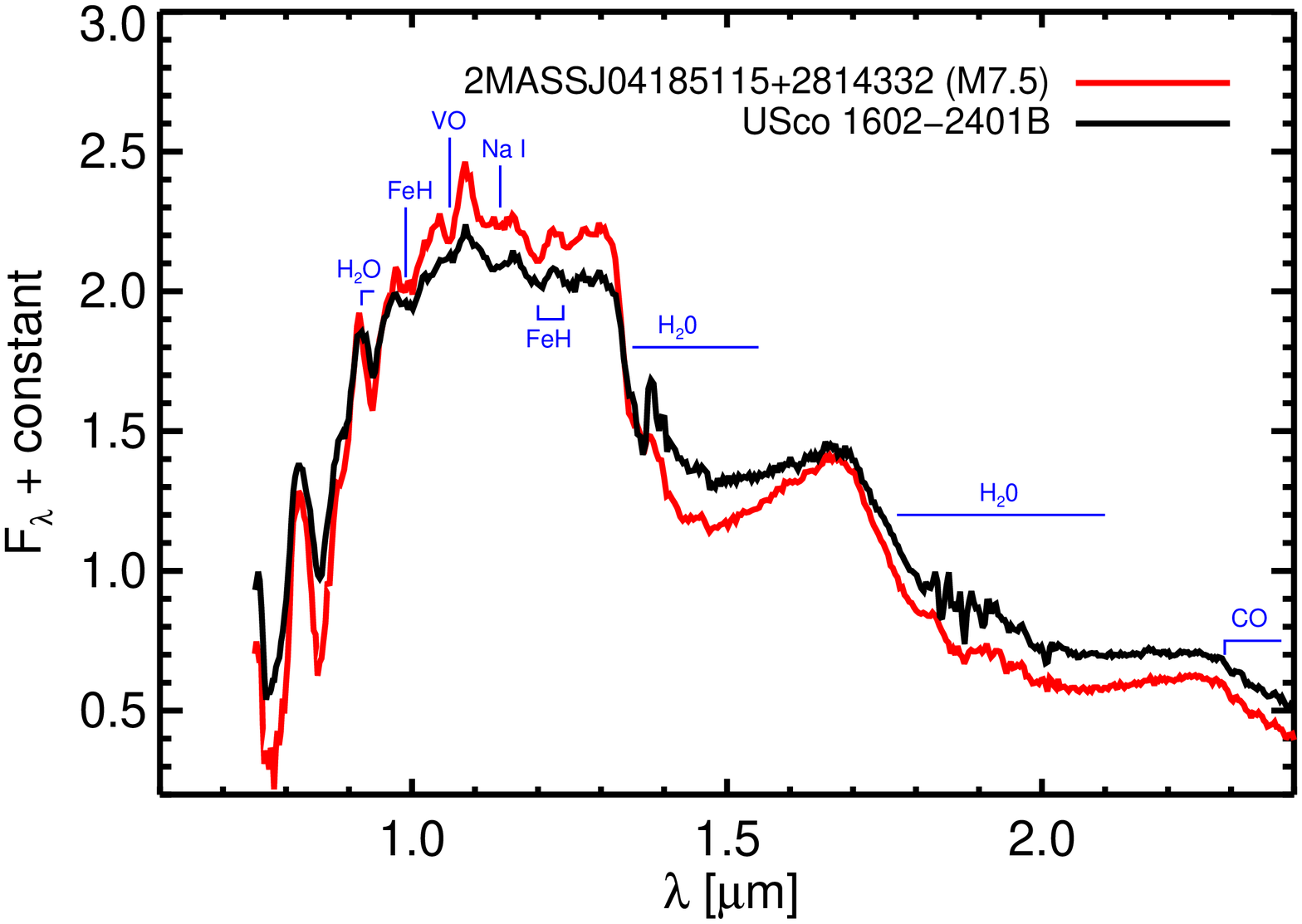}
    \caption{USco\,1602$-$2401B has a projected separation of
      1000\,$\pm$\,14\,AU and a spectral type of M7.5
      ($\approx$\,41\,\Mjup). \emph{LEFT} -- The finder chart (UKIDSS
      $K$ and width of 30\arcsec) with the letter B identifying the
      companion. North is \emph{up} and East is
      \emph{left}. \emph{RIGHT} -- IRTF SpeX spectrum of
      USco\,1602$-$2401B compared to the young (1--2\,Myr) M7.5 in
      Taurus
      \citep[2MASS~J04185115+2814332;][]{2007AJ....134..411M}. \label{fig:f6}}
  \end{center}
\end{figure}

\begin{figure}
  \begin{center}
    \includegraphics[height=0.4\textwidth]{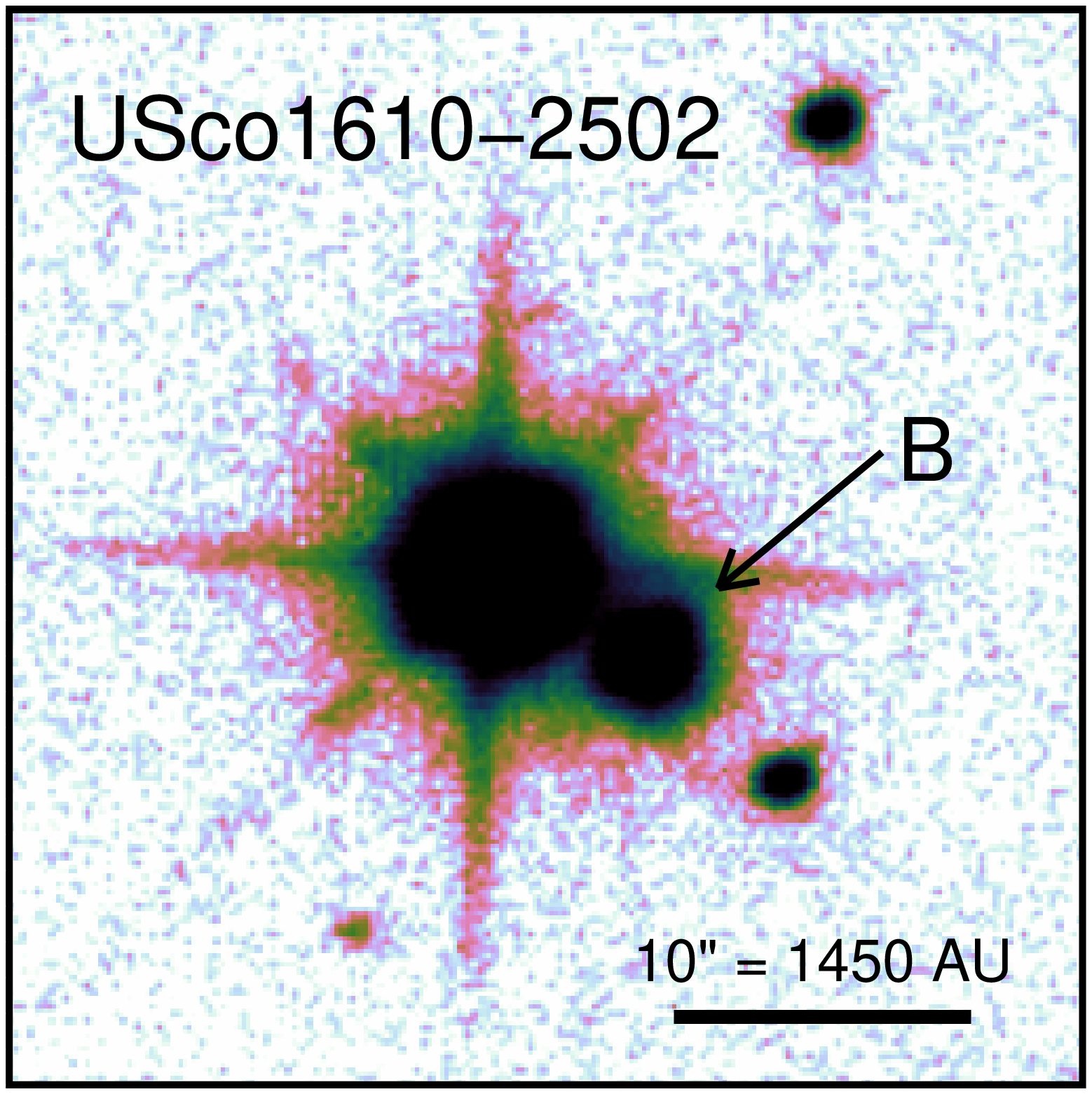}
    \includegraphics[height=0.4\textwidth]{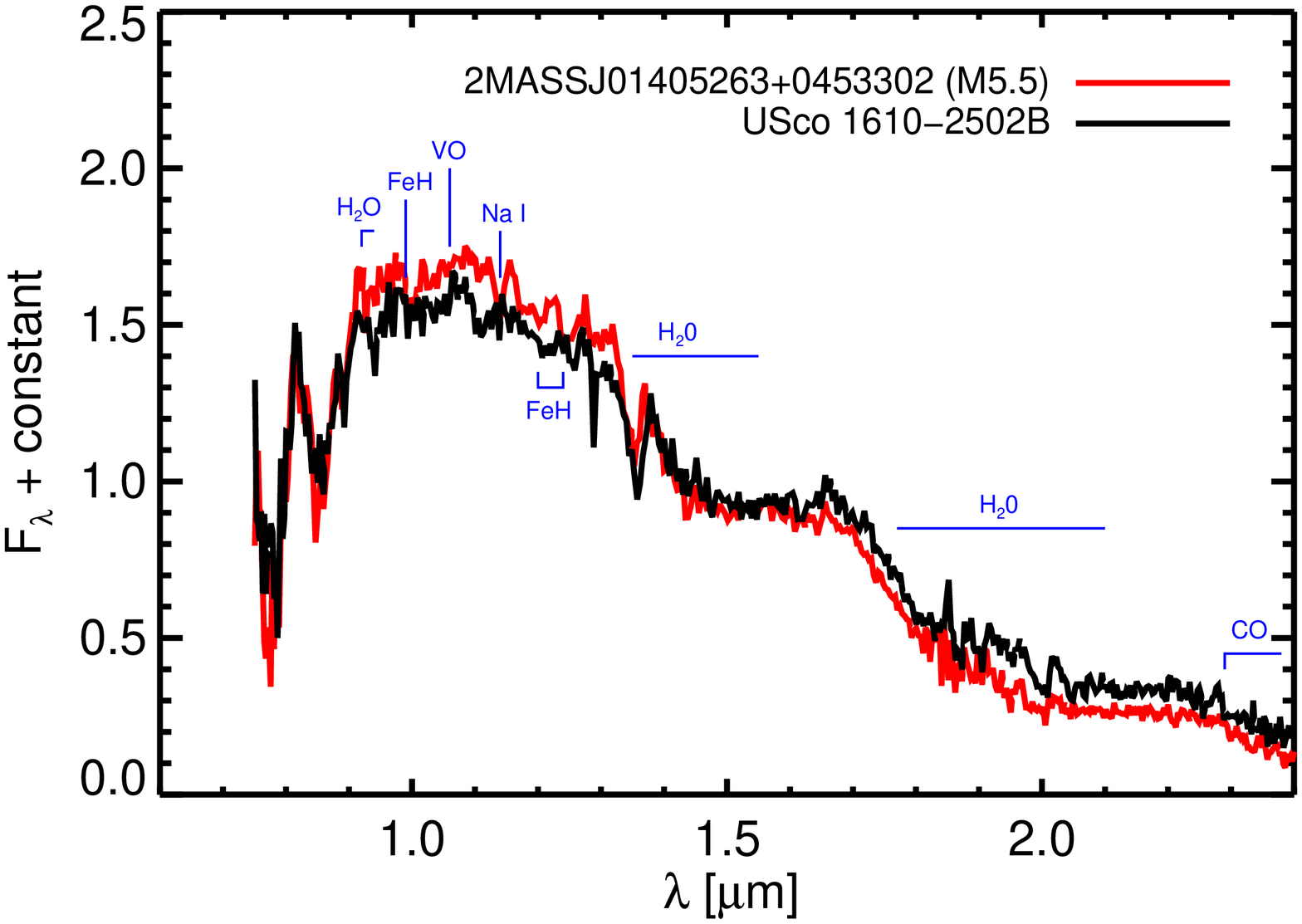}
    \caption{USco\,1610$-$2502B has a projected separation of
      732\,$\pm$\,76\,AU and a spectral type of M5.5
      ($\approx$\,0.13\,\Msun). \emph{LEFT} -- The finder chart
      (UKIDSS $K$ and width of 30\arcsec) with the letter B
      identifying the companion. North is \emph{up} and East is
      \emph{left}. \emph{RIGHT} - IRTF SpeX spectrum of
      USco\,1610$-$2502 compared to the field M5.5
      \citep[2MASS~J04185115+0453302;][]{2010ApJS..190..100K}. \label{fig:f7}}
  \end{center}
\end{figure}

Our spectroscopic observations of 13 candidates yielded five new
companions. Four are new substellar companions with spectral types M9
(HIP\,77900B), M9 (USco\,1610$-$1913B), M8.5 (USco\,1612$-$1800B), and
M7.5 (USco\,1602$-$2401B). One is a low-mass stellar companion with
spectral type M5.5 (USco\,1610$-$2502B). Figures~\ref{fig:f3},
\ref{fig:f4}, \ref{fig:f5}, \ref{fig:f6}, and \ref{fig:f7} show the
UKIDSS $K$ image and the reduced spectrum with the closest matching
published spectrum for each of the companions. We describe them in
more detail in the following sections.

\subsubsection{HIP 77900B}

HIP\,77900 is a B6V star \citep{1967ApJ...147.1003G} with a
model-dependent mass of 3.8\,$^{+0.7}_{-0.5}$\,\Msun\ at 5\,Myr
(3.8\,$^{+0.8}_{-0.5}$\,\Msun\ at 10\,Myr). Its membership in Upper
Sco was first determined using both {\it Hipparcos} proper motions and
parallax by \citet{1999AJ....117..354D} and later by
\citet{2011MNRAS.416.3108R} with the addition of radial velocities.

Our new companion, HIP\,77900B, has a projected separation of
3200\,$\pm$\,300\,AU and a spectral type of M9\,$\pm$\,0.5. The mass
ratio between the primary for an age of 5\,Myr, is
$q$\,=\,0.005$\pm$0.002 ($q$\,=\,0.005$\pm$0.003 at 10\,Myr). Mass
ratios this small are rare but have been observed, such as for the
HR\,7329 system \citep[$q$\,$\sim$\,0.01;][]{2000ApJ...541..390L} and
HD\,1160 \citep[$q$\,$\sim$\,0.014 for the B
  component;][]{2012ApJ...750...53N}. Although we have not determined
if HIP\,77900B is co-moving and bound to HIP\,77900, its triangular
H-band continuum confirms its youth and therefore membership in Upper
Sco. Furthermore, \citet{2008ApJ...686L.111K} found that binaries can
be distinguished from chance alignments in Upper Sco at separations
$\lesssim$\,75\arcsec, down to primaries with
M\,$\gtrsim$\,0.3\,\Msun. The 22\,\arcsec\ separation of HIP\,77900B
suggests that it is truly associated with HIP\,77900.

\subsubsection{USco\,1610$-$1913B}

\citet{2002AJ....124..404P} identified USco\,J161031.9$-$191305
(hereafter USco\,1610$-$1913) as a member of Upper Sco based on
lithium absorption ($EW$[Li]\,=\,0.55\,\AA) and weak H$\alpha$
emission ($EW$[H$\alpha$]\,=\,-2.3\,\AA). Furthermore they determined
a spectral type of K7 and $A_{V}$\,=\,1.1\,mag. At 5\,Myr,
USco\,J1610$-$1913 has a model-dependent mass of
0.88$^{+0.14}_{-0.17}$\,\Msun\ (0.87$^{+0.11}_{-0.18}$\,\Msun\ at
10\,Myr).

\citet{2009ApJ...703.1511K} determined that 2MASS\,16103232-1913085
(hereafter USco\,1610$-$1913B) and USco\,J1610$-$1913 are co-moving
with a projected separation of 840\,$\pm$\,90\,AU. Using the flux
ratio with the primary, \citet{2009ApJ...703.1511K} estimate a mass of
34\,\Mjup. The $K$ magnitude ($K$\,=\,12.74\,$\pm$\,0.002\,mag) also
suggests that the mass, assuming an age of 5\,Myr, is
$\approx$\,34\,\Mjup\ and the spectral type is M7.8, using the methods
described in Section~\ref{sec:physicalproperties}. However we
spectroscopically confirmed a later spectral type of
M9\,$\pm$\,0.5. Assuming an age of 5\,Myr, the model-dependent mass is
19$^{+7}_{-4}$\,\Mjup\ (20$^{+7}_{-3}$\,\Mjup\ at 10\,Myr),
significantly lower than predicted from its absolute magnitude.

USco\,1610$-$1913 has a small $K_{s}-W4$ excess
(1.1\,$\pm$\,0.2\,mag), which is marginally consistent with the presence
of a disk. However, there is no excess at 4.5$\mu$m, 8.0$\mu$m or
24$\mu$m \citep{2012ApJ...758...31L}, and thus there is likely no
disk.

\subsubsection{USco\,1612$-$1800B}

\citet{2002AJ....124..404P} confirmed USco\,161248.9$-$180052
(hereafter USco\,1612$-$1800) as a member of Upper Sco based on
lithium absorption ($EW$[Li]\,=0.52\,\AA) and H$\alpha$ emission
($EW$[H$\alpha$]\,=-3.8\,\AA). They also determined the spectral type
to be M3 with $A_{V}$\,=\,1.4\,mag. Assuming an age of 5\,Myr, the
model-dependent mass is
0.36$^{+0.14}_{-0.12}$\,\Msun\ (0.36$^{+0.14}_{-0.15}$\,\Msun\ at
10\,Myr).

USco\,1612$-$1800B has the smallest projected separation of our new
companions (430\,$\pm$\,40\,AU). With a spectral type of M8.5,
USco\,1612$-$1800B has a mass of 23$^{+12}_{-6}$\,\Mjup, assuming an
age of 5\,Myr (26$^{+16}_{-7}$\,\Mjup\ at 10\,Myr). Although it has
not been confirmed as a comoving companion to USco\,1612$-$1800, the
spectrum shows signatures of youth consistent with its membership in
Upper Sco and the small separation ($\sim$\,3\arcsec) further implies
that it is likely a bound system \citep{2008ApJ...686L.111K}.

Although \citet{2012ApJ...758...31L} concluded that USco\,1612$-$1800
has $W4$ excess and therefore hosts a debris/evolved transitional
disk, USco\,1612$-$1800 and USco\,1612$-$1800B are separated by about
3\arcsec, and thus unresolved by {\it WISE} (whose PSF FWHM range from
6--12\arcsec). The 2MASS\,$-${\it WISE} color is high,
$K_{s}-W4$\,=\,2.24\,$\pm$\,0.26\,mag, but this represents the
integrated light of the binary. We must deblend the $W4$ magnitude in
order to conclude whether the primary or secondary has a disk. If we
assume a typical $K_{S}-W4$ color for Upper Sco late-M stars with
disks,
\citep[$K_{S}-W4$\,$\approx$\,4--6\,mag;][]{2012ApJ...758...31L} we
can estimate the $W4$ flux of the secondary to be
$W4$\,=\,7.2--9.2\,mag. We also know that the secondary cannot be
brighter than the integrated-light $W4$ magnitude of 8.1\,mag. If
there is a disk around the secondary, the resulting $K_{S}-W4$ for the
primary is $K_{S}-W4$\,$\lesssim$\,1.75\,mag. This color still
suggests that USco\,1612$-$1800 may host a debris/evolved transitional
disk but is also consistent with no disk. We conclude that there is
likely a debris/evolved transitional disk in this system but cannot
conclude if it resides around the primary or secondary.

\subsubsection{USco\,1602$-$2401B}

2MASS\,J16025123$-$2401574 is a K4 member of Upper Sco (hereafter
USco\,1602$-$2401), identified as an X-ray source and
spectroscopically confirmed by \citet{1999PhDT........89K}. At 5\,Myr,
the model-dependent mass is
1.34$^{+0.12}_{-0.13}$\,\Msun\ (1.18$^{+0.06}_{-0.07}$\,\Msun\ at
10\,Myr).

\citet{2009ApJ...703.1511K} confirmed that 2MASS\,J160251.16-240150.2
(hereafter USco\,1602$-$2401B) is a co-moving companion to
USco\,1602$-$2401 with a projected separation of
1000\,$\pm$\,140\,AU. USco\,1602$-$2401B has a mass at 5\,Myr is
41$^{+20}_{-13}$\,\Mjup\ (47$^{+20}_{-18}$\,\Mjup\ at 10\,Myr). This
mass is significantly lower than the previous estimate,
$\sim$\,0.11\,\Msun, based on the flux ratio relative to the primary
\citep{2009ApJ...703.1511K}. 

The $K_{S}-W4$ color ($K_{s}-W4$\,$\le$\,0.98\,mag) indicates that at
most there is weak excess from any possible disk. However,
\citet{2012ApJ...758...31L} find a 24\,$\mu$m excess and conclude that
USco\,1602$-$2401 does have a debris/evolved transitional disk.

\subsubsection{USco\,1610$-$2502B}

\citet{1998AA...333..619P} identified USco\,161019.18$-$250230.1
(hereafter USco\,1610$-$2502) as an X-ray source and determined its
membership in Upper Sco based on lithium absorption
($EW[Li]$\,=0.52\,\AA), $H\alpha$ emission
($EW[H\alpha]$\,=\,$-$0.75\.\AA). They also determined the spectral
type to be M1. We find a model-dependent mass of
0.70$^{+0.20}_{-0.20}$\,\Msun\ assuming an age of 5\,Myr
(0.70$^{+0.18}_{-0.17}$\,\Msun\ at 10\,Myr).

Our new companion, USco\,1610$-$2502B, is a confirmed proper motion
companion to USco\,1610$-$2502 \citep{2009ApJ...703.1511K}. It has a
projected separation of 730\,$\pm$\,80\,AU and a model-dependent mass
at 5\,Myr of
0.10$^{+0.08}_{-0.05}$\,\Msun\ (0.09$^{+0.07}_{-0.04}$\,\Msun\ at
10\,Myr). Although its spectrum does not have the obvious triangular
H-band continuum that distinguishes young ultracool dwarfs from older
field objects, this feature is less pronounced for mid-M spectral
types.

We find no evidence for a disk around USco\,1610$-$2502 according to
its 2MASS\,$-${\it WISE} color, $K_{s}-W4 <
1.2$\,mag. \citet{2012ApJ...758...31L} also conclude that USco\,1610$-$2502
has no disk because they detect no excesses at 4.5\,$\mu$m, 8.0\,$\mu$m,
or 24\,$\mu$m. However, there is evidence for a disk around
USco\,1610$-$2502B according to its 2MASS\,$-${\it WISE} colors, $K_{s}-W4
= 4.9 \pm 0.2$\,mag and $K_{s}-W3 = 3.09 \pm 0.06$\,mag.

\subsection{Background Objects}

We identified a total of nine background objects
(Table~\ref{table:tab1}) and compared the NIR spectra with the
IRTF/SpeX Prism Library to visually classify them. We also fit these
objects with a reddened blackbody to determine if they were reddened
early-type stars, i.e. background stars. Two of our background objects
had flat spectra that could not be reproduced with a reddened
blackbody and thus may be galaxies.


\section{Discussion} \label{sec:discussion}

\begin{figure}
  \begin{center}
    \includegraphics[angle=90, width=0.9\textwidth]{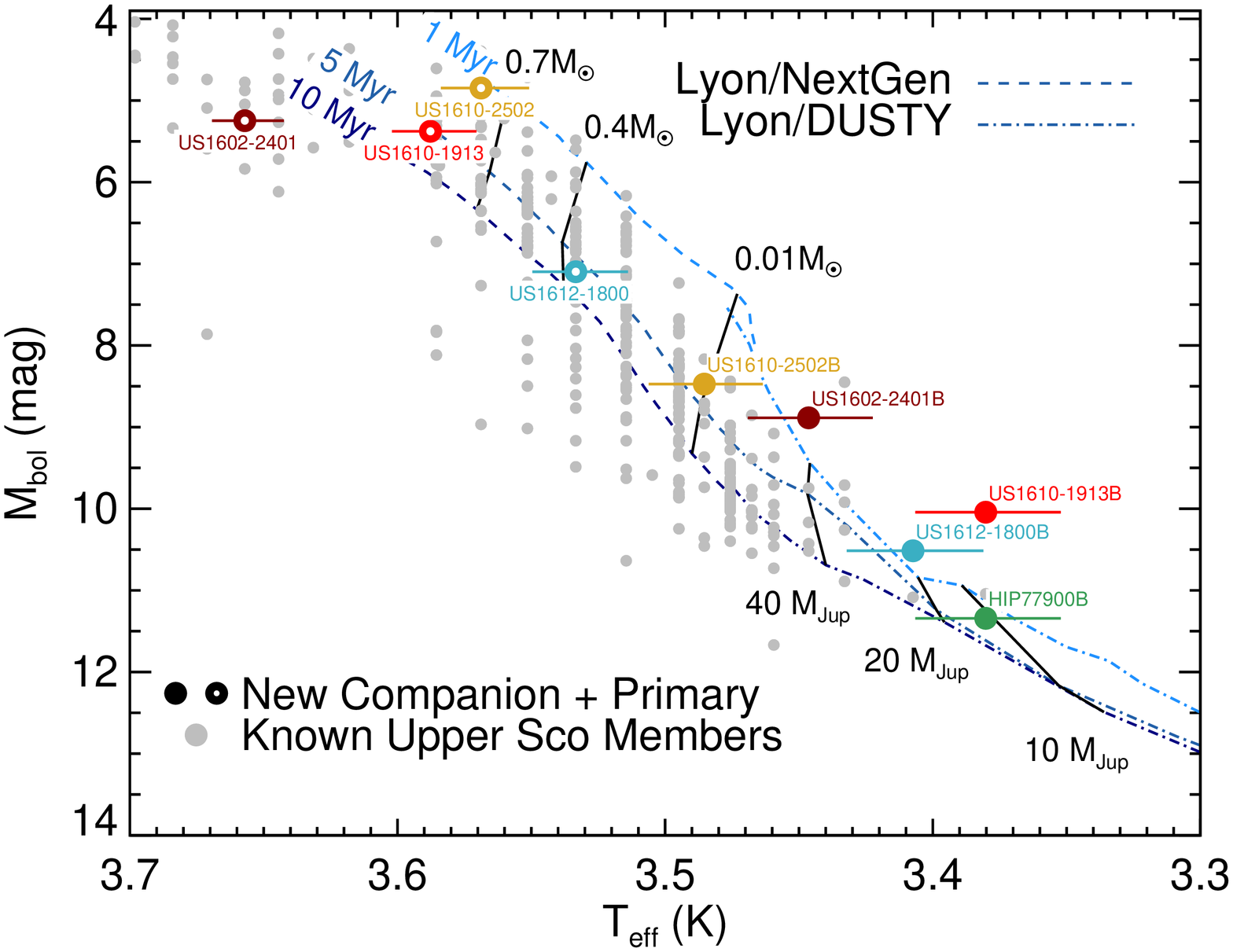}
    \caption{The bolometric luminosity and temperature of our new
      companions and their primaries compared to the model grids are
      the \emph{solid} and \emph{open circles} for the companions and
      the primaries, respectively \citep{2000ApJ...542..464C,
        1998AA...337..403B}. We omitted HIP~77900 because it has a
      very high mass and $T_{eff}$ compared to the rest of the
      sample. Each system is represented by a different color. The
      \emph{gray circles} are the Upper Sco free-floating members from
      \citet{2007ApJ...662..413K}. The three \emph{dashed} colored
      lines are isochrones for 1\,Myr (\emph{light blue}), 5\,Myr
      (\emph{blue}), and 10\,Myr (\emph{dark blue}). The \emph{solid
        black lines} trace out the evolution for an object with a
      given mass. The error bars for the uncertainty in bolometric
      magnitude, 0.13\,mag, are smaller than the data
      points. \label{fig:f8}}
  \end{center}
\end{figure}

In Figure~\ref{fig:f8} we show an HR diagram with our newly identified
companions, the free-floating Upper Sco members, and the
\citet{1998AA...337..403B} and \citet{2000ApJ...542..464C}
models. The Upper Sco free-floating members in this diagram are taken
from \citet{2007ApJ...662..413K}, \citet{2008ApJ...688..377S}, and
\citet{2008MNRAS.383.1385L}. Our substellar companions as a group have
systematically higher bolometric magnitudes than the observed cluster
sequence (Figure~\ref{fig:f1} and Figure~\ref{fig:f8}). Within the
uncertainties in effective temperature, USco\,1610$-$1913 and
USco\,1602$-$2401B clearly have a higher bolometric luminosities than
both the known members and the models. We note that there is one other
previously known M8 member, 2MASS~J162243.85$-$195105.7, that is also
overluminous compared to the observed cluster sequence and the models
though it may be a spectroscopic binary \citep{2012ApJ...745...56D}.

Furthermore, the primaries of our substellar companions as a group are
not overluminous compared to the models (Figure~\ref{fig:f8}),
suggesting that the primary and companion may not be coeval in all
cases. Although higher mass ($\sim$\,0.1--1.5\,\Msun) young binary
systems appear coeval \citep{2009ApJ...704..531K}, our results suggest
that this may not be true for lower mass companions.

Finally, USco\,1610$-$1913B is also more luminous ($\sim$\,1.5\,mag)
than HIP\,77900B despite having the same spectral type (M9). The
primary, USco\,1610$-$1913, is not overluminous compared to the models
although it has marginal evidence for a disk
(Section~\ref{sec:results}) which could contaminate the $K$-band
magnitude, and hence the calculated bolometric luminosity.

One possible reason for the overluminosity of young companions
compared to the models could be different accretion histories
\citep{2012ApJ...756..118B}. Young stars may have strong episodic
accretion which will increase the star's radius and thus the
luminosity. For example, \citet{2011ApJ...743..148B} found that the
planetary-mass companion GSC~06214-00210b in Upper Sco likely has
strong accretion from a circumplanetary disk. Strong accretion could
also explain the discrepancy of 1.5\,mag between the luminosities of our
two new M9 companions, HIP\,77900B and USco\,1610$-$1913B.

The wide projected separations of our new companions are difficult to
explain as either the massive-end of gas-giant planet formation or the
low-mass tail of binary-star formation. Binary star systems can have
very wide separations up to several thousand AU
\citep[e.g.][]{1991AA...248..485D} but it is still unclear whether
wide binaries can form with such an extreme separation
($\sim$\,3100\,AU) and mass ratio ($\sim$\,0.005) as the HIP\,77900AB
system. However, even if they formed from a protoplanetary disk,
planet formation has only been modeled to a few hundred AU
\citep[e.g.][]{2010ApJ...710.1375K,2010MNRAS.406.2279M}. Thus, whether
protoplanetary disks can also create such wide planetary-mass
companions remains uncertain.


\section{Conclusions} \label{sec:conclusion}

We have used PS1 and UKIDSS photometry to search for wide
($\approx$\,400--4000\,AU) planetary-mass companions in Upper Sco down
to $\sim$\,5\,\Mjup. We use a selection method that combines
traditional color-selection with SED fitting. Our method significantly
decreases the number of reddened background stars that contaminate our
sample compared with a solely color-selected sample.

We obtained followup low-resolution NIR spectroscopy of several
candidates and discovered two new companions and confirmed three other
companions. Four are very low mass substellar companions (spectral
type M7.5-M9, mass $\approx$\,15-60\,\Mjup) and one is a low-mass star
(spectral type M5, mass $\sim$\,0.1\,\Msun). The most extreme object
is HIP\,77900B because of its very wide projected separation
(3200\,$\pm$\,300\,AU) and very small mass ratio
($q$\,$\approx$\,0.005). 

Altogether, we have spectroscopically confirmed 4 wide substellar
companions out of our search around 673 Upper Sco members. Our results
indicate a frequency for wide (400-4000\,AU) substellar companions
down to 5\,\Mjup\ of $\sim$\,0.6\,$\pm$\,0.3\%. The wide projected
separations for all of our companions are difficult to explain as
either the massive-end of planetary formation or the low-mass tail of
binary-star formation.

In addition, two of our companions (USco\,1610$-$1913B and
HIP\,77900B) present another puzzle, because they have the same
spectral type but luminosities that differ by 1.5\,mag. Altogether,
our companions suggest that young substellar companions, but not
necessarily their respective primary stars, are overluminous compared
to the models and the observed cluster sequence. As a result, our new
companions do not all appear coeval with their primary stars on an HR
diagram, in contrast to results for young higher mass binary systems
\citep{2009ApJ...704..531K}.

Regardless of the formation scenario of these companions, we can use
them as young spectral benchmarks to constrain evolutionary
models. These new companions provide us with a unique glimpse into the
early life of brown dwarfs. Further discoveries will improve our
constraints on models and our understanding of planet/brown dwarf
formation, the typical properties of these young systems, and their
likely evolution.


\acknowledgments 

KMA's research was supported by the National Science Foundation under
Grant No. 0822443. Any opinion, findings, and conclusions or
recommendations expressed in this material are those of the authors'
and do not necessarily reflect the views of the National Science
Foundation. ALK was supported by a Clay Fellowship and by NASA through
Hubble Fellowship grant 51257.01, awarded by STScI. KMA and MCL were
also supported by NSF Grant No. AST09-09222. We thank Brendan
P. Bowler for obtaining some of the IRTF/SpeX data. We used data
products from the UKIRT Infrared Deep Survey (UKIDSS). The
Pan-STARRS~1 Surveys (PS1) have been made possible through
contributions of the Institute for Astronomy, the University of
Hawaii, the Pan-STARRS Project Office, the Max-Planck Society and its
participating institutes, the Max Planck Institute for Astronomy,
Heidelberg and the Max Planck Institute for Extraterrestrial Physics,
Garching, The Johns Hopkins University, Durham University, the
University of Edinburgh, Queen's University Belfast, the
Harvard-Smithsonian Center for Astrophysics, the Las Cumbres
Observatory Global Telescope Network Incorporated, the National
Central University of Taiwan, the Space Telescope Science Institute,
and the National Aeronautics and Space Administration under Grant
No. NNX08AR22G issued through the Planetary Science Division of the
NASA Science Mission Directorate. We also use data products from the
Two Micron All Sky Survey, which is a joint project of the University
of Massachusetts and the Infrared Processing and Analysis
Center/California Institute of Technology, funded by the National
Aeronautics and Space Administration and the National Science
Foundation. In addition, we use of data products from the {\it
  Wide-field Infrared Survey Explorer}, which is a joint project of
the University of California, Los Angeles, and the Jet Propulsion
Laboratory/California Institute of Technology, funded by the National
Aeronautics and Space Administration. This research has also
benefitted from the M, L, and T dwarf compendium housed at
DwarfArchives.org and maintained by Chris Gelino, Davy Kirkpatrick,
and Adam Burgasser. We also use the IRTF Prism Spectral Library housed
at \url{http://pono.ucsd.edu/\textasciitilde
  adam/browndwarfs/spexprism} and maintained by Adam Burgasser. We
have also made use of TOPCAT, an interactive tool for manipulating and
merging tabular data. Finally, mahalo nui loa to the kama'\={a}ina of
Hawaii'i for allowing us to operate telescopes on Mauna Kea. We wish
to acknowledge the very significant culutural role Mauna Kea has
within the indigenous Hawaiian community and that we are very
fortunate to be able to conduct observations.

Facilities: IRTF (SpeX), UKIRT (UKIDSS), Pan-STARRS~1 (3\,$\pi$\,Survey).


\appendix
\section{SED Templates} \label{appendix:sed_templates}

We characterized the PS1 and UKIDSS color-magnitude relations for cool
and ultracool dwarf stars (spectral type M0--L0) using open cluster
dwarfs from Praesepe and Coma Berenices, young open cluster dwarfs
from Upper Sco, and field dwarfs with parallaxes. In order to increase
the flexibility of our program to fit to a variety of spectral energy
distributions (SEDs), we also use field dwarfs with optical+NIR SEDs
but without parallaxes and separate our fitting procedure into two
parts: spectral type and photometric distance. We then use the
resulting templates to determine the spectral types and the
photometric distances of our candidates.


\subsection{Average Templates from Clusters}

\subsubsection{Open Cluster Dwarfs}
Our template Praesepe and Coma Berenices members are selected from
\citet{2007AJ....134.2340K}. Praesepe ($\sim$\,600\,Myr) is located at
a distance of 170\,pc \citep{1995AAS..109...29H} and Coma Berenices
($\sim$\,400\,Myr) is at 90\,pc \citep{2006MNRAS.365..447C}. We use
the Praesepe and Coma Berenices members with $\ge$95$\%$ membership
probability, spectral type $\ge$ M0, and good quality photometry (in
\ips, \zps, or \yps) in the PS1 3$\pi$\ survey DVO catalog. After
applying our data quality cuts (see Section~\ref{sec:data}), in total
we had 506 cluster dwarfs (spectral type M0$-$M5) with a mixture of
PS1 magnitudes, \ips, \zps, or \yps, and UKIDSS magnitudes,
$ZYJHK$.

In order to create the SED templates for the Praesepe and Coma
Berenices members as a function of spectral type, we computed the
weighted average for the absolute magnitude for each half spectral
subclass (i.e. M0, M0.5, etc). We remove the binary sequence by
selecting the stars which are brighter than a simple quadratic line
which runs above most of the data. After accounting for the binary
sequence in this fashion, we compute a 0.48~mag scatter in the
absolute magnitudes for a given spectral type. This dispersion is
consistent with previous studies of the absolute magnitudes of
M~dwarfs \citep[e.g.][]{2010AJ....139.2679B,2012ApJS..201...19D}. The
final SEDs are tabulated in Table~\ref{table:tab3}.

\subsubsection{Upper Sco Members}
We also created SED templates from known Upper Sco members spanning
spectral type M0 to L2. Earlier type objects were saturated in PS1
and/or UKIDSS. Just as for the open cluster dwarfs, we selected UKIDSS
and PS1 photometry with our same data quality restrictions. The final
SED templates are the weighted average of the absolute magnitudes for
each half spectral subclass (i.e. M0, M0.5, etc). Note that the large
errors reflect the known large spread in absolute magnitude seen in
the known members of Upper Sco\citep[e.g.][]{2008MNRAS.383.1385L}. In
total, we used 404 Upper Sco members stars to create the final SEDs
although a different numbers of primaries are used for each SED
average magnitude because of the variable coverage by both UKIDSS and
PS1. The final SEDs are tabulated in Table~\ref{table:tab4}.


\subsection{Templates of Individual Field M and L Dwarfs}

The M, L $\&$ T dwarfs are from taken from the
\citet{2009AJ....137....1F} and \citet{2010ApJ...710.1627L} ultracool
dwarf catalogs and DwarfArchives.org.  By separating our fitting
routine into two parts (spectral typing and distance determination),
we can use significantly more templates (a selection from
$\approx$\,1000 known field dwarfs instead of from only about 100 that
have parallaxes). This method also better encompasses the SED
variation within a given spectral type. The average properties do not
fully convey their diversity
\citep[e.g.][]{2001ApJ...548..908L}. Furthermore, we use the SEDs of
individual field dwarfs, rather than constructing average SEDs for
each spectral type, as for cluster dwarfs, as the final SED
templates. Thus, the uncertainty in the templates are just the actual
measurement errors in the PS1 and UKIDSS magnitudes. In our analysis,
we also exclude dwarfs which are known binaries or close binaries (81)
and without at least 3 good quality measurements from UKIDSS or PS1
(see Section~\ref{sec:data}). Therefore, the final set of dwarfs used
to determine the spectral type has 115 dwarfs. For this work, only
objects with spectral type M9--L0 are tabulated in
Table~\ref{table:tab5}. Although our fit includes the entire library
of field dwarfs we do not expect to find any late~L or T~dwarfs in
this work. The colors and details of PS1 colors for L/T dwarfs is
reserved for a later paper where those details will be
discussed. L/T~dwarf colors for UKIDSS are already reported in other
papers \citep[e.g.][]{2001ApJ...548..908L,2006MNRAS.367..454H}.

Finally, in order to determine the absolute magnitude as a function of
spectral type, we use the 30 ultracool dwarfs (spectral type M7 to L0)
with parallaxes and good quality measurements (see
Section~\ref{sec:data}) in PS1 and/or UKIDSS. Therefore, we only
use these 30 dwarfs with parallaxes to determine absolute magnitude
from spectral type but use the full library of field dwarfs (115) to
determine the spectral type from the SED.


\subsection{Final Library of SED Templates}
\begin{figure}
  \begin{center}    
    \includegraphics[width=.45\textwidth]{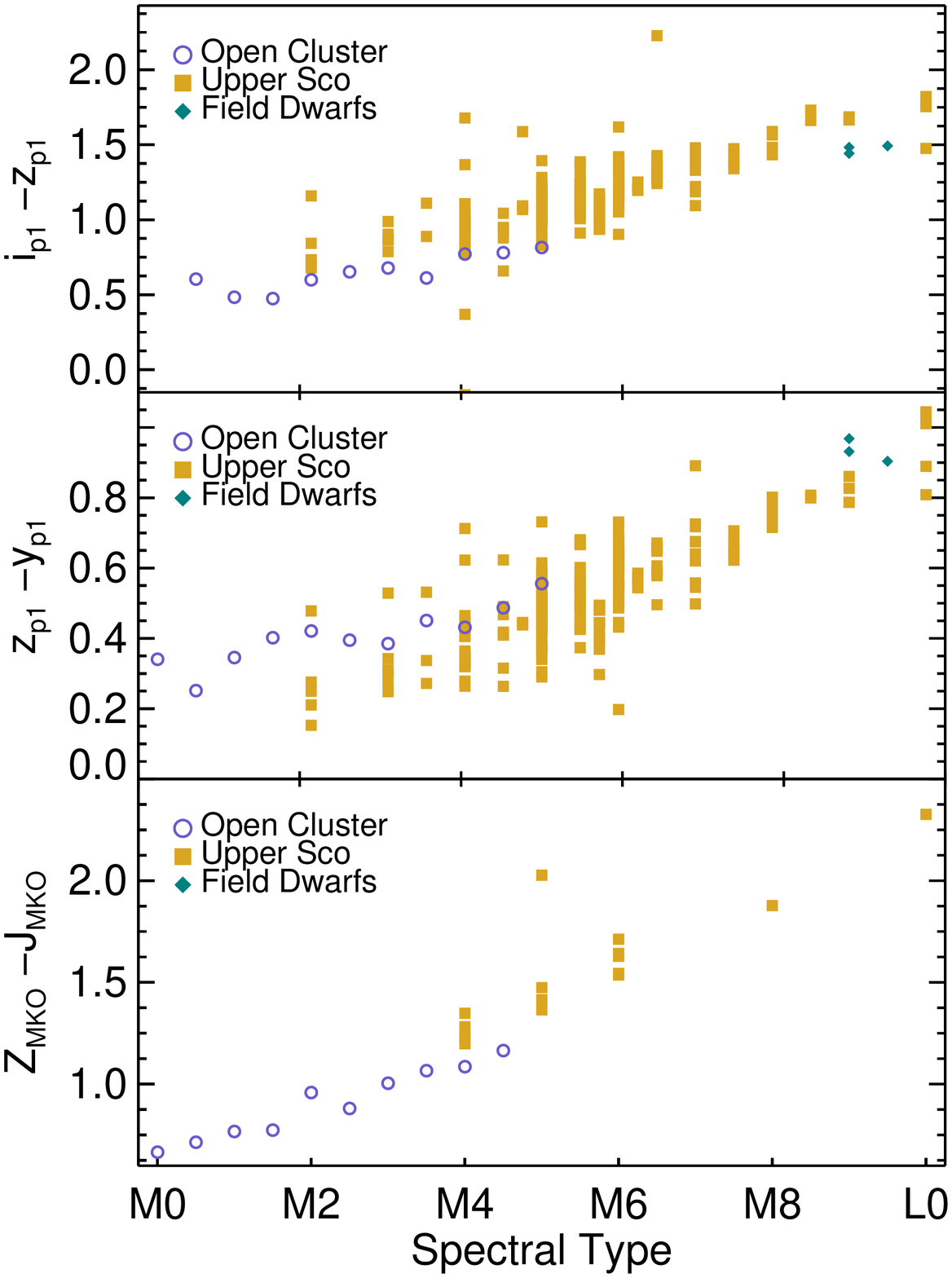}
    \includegraphics[width=.45\textwidth]{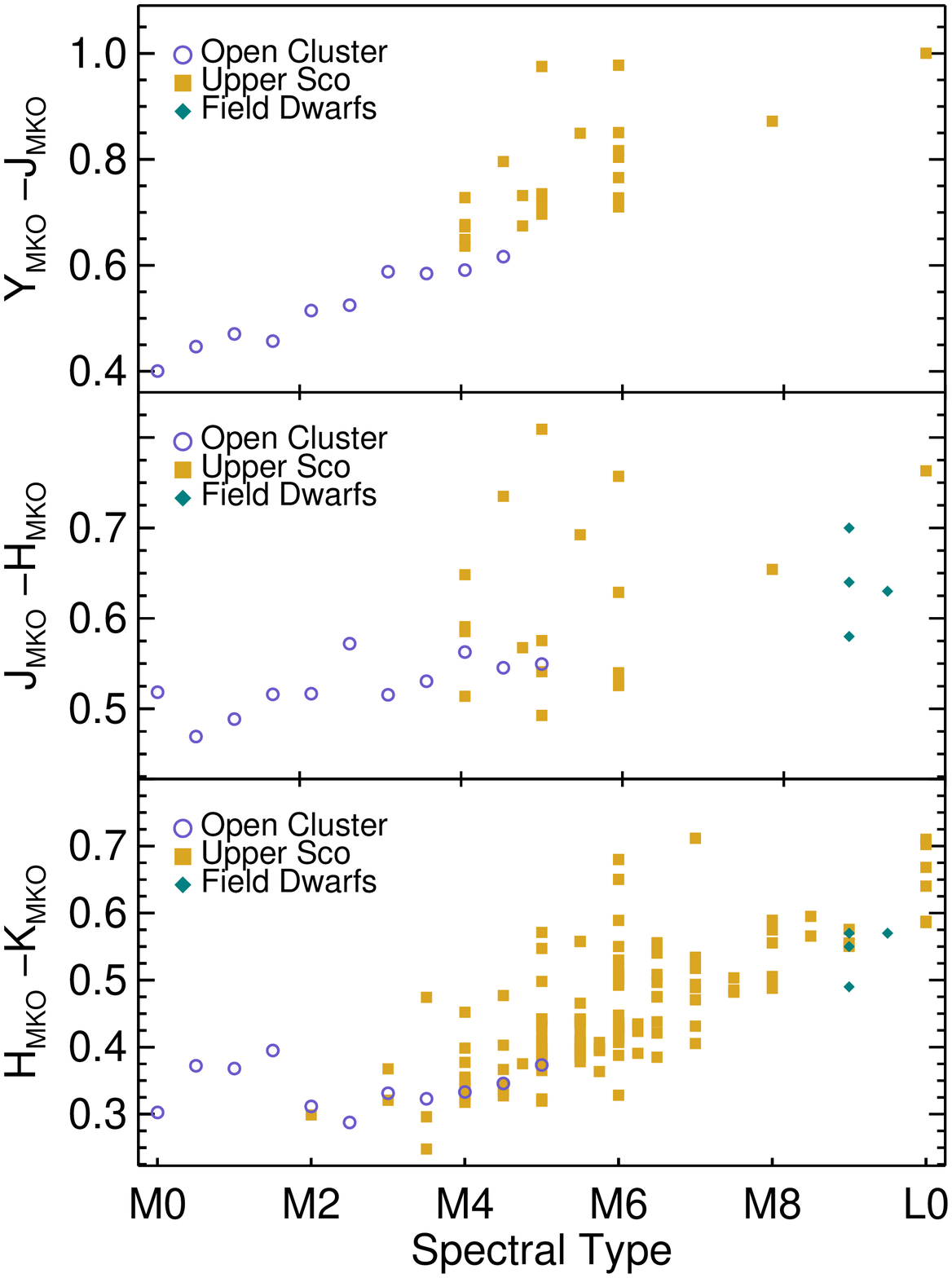}
   \caption{The optical and near-IR colors with \ips, \zps, \yps, and
     the UKIRT $ZYJHK$ as a function of the spectral type for all the
     SED templates used in our spectral type routine. These templates
     all are constructed from individual objects with magnitude errors
     $\le$\,0.2\,mag and have detections in at least 3 filters
     (including one in the UKIDSS filters). The M~dwarf templates
     constructed from averages from Praesepe and Coma Berenices
     members are the \emph{purple open circles}. The individual SEDs from
     known Upper Sco M--L0~dwarf members are the \emph{gold squares}
     and show the large spread in color in Upper Sco. The field dwarf
     SEDs are the \emph{teal diamonds}. \label{fig:f9}}
  \end{center}
\end{figure}

The final SEDs used to perform the $\chi^{2}$ fit for spectral type
have at least three of the following: \ips, \zps, \yps, $Z$, $Y$, $J$,
$H$, or $K$. There are a total of 133 templates covering the spectral
type range M0--L2, where 10 are averages from the cluster dwarfs
(Table~\ref{table:tab3}), 8 are from Upper Sco primary stars
(Table~\ref{table:tab4}) and 115 are field dwarfs
(Table~\ref{table:tab5}). Upper Sco
free-floating cluster members fill the gap at M6--M8 (between the open
cluster members in Praesepe and Coma Berenices and field dwarfs) where
the majority of our candidates may be; Brown dwarfs in Upper Sco are
expected have spectral types greater than M7. We show the optical-NIR
colors as a function of spectral type for final SED templates in
Figure~\ref{fig:f9}. The observed scatter in color as a function of
spectral type highlights the importance of using several SED templates
for each spectral type. Finally, we use the weighted average cluster
dwarf SEDs and only 30 field dwarf SEDs (with parallaxes) for the
absolute magnitude-spectral type relation which covers the spectral
type range M0--L8 with a few gaps. In our analysis, this relationship
is only used to convert from the fitted spectral type to absolute
magnitude as a check on whether the photometric distance is consistent
with Upper Sco membership. For example, objects which fit
early-M~dwarfs will generally be farther (taking into account the
higher luminosities observed for young stars) than expected and, thus,
can be flagged as background stars.


\clearpage 

\bibliographystyle{apj}


\clearpage




\clearpage

\begin{landscape}
\begin{deluxetable}{cccccccccc}
  \tablecaption{IRTF Observations \label{table:tab1}}
  \tabletypesize{\tiny}
  \tablewidth{0pt}
  \tablehead{
    \colhead{Name} &
    \colhead{RA} &
    \colhead{Dec} &
    \colhead{Date} &
    \colhead{$T_{exp}$} &
    \colhead{Spectral Type} &
    \colhead{Separation} & 
    \colhead{PA} &
    \colhead{$K$} &
    \colhead{A0V Standard}
    \\ 
    \colhead{ } &
    \colhead{(J2000)} &
    \colhead{(J2000)} &
    \colhead{(UT)} &
    \colhead{(sec) } &
    \colhead{ } &
    \colhead{(arcsec)} &
    \colhead{(deg)} &
    \colhead{(mag)} &
    \colhead{ }
  }
  \startdata
  \multicolumn{10}{c}{New Candidates} \\
  & & & &                      & & & & & \\
  USco\,1602$-$2401B & 16:02:51.17 & -24:01:50.45      & 2011 June 19 & 960 & M7.5$\pm$0.5 & 7.0 & 352.1$^{\circ}$   & 11.60$\pm$0.05\tablenotemark{a} & HD\,145127 \\ 
  \ldots & \ldots & \ldots & 2013 April 16 & 360 & \ldots & \ldots & \ldots & \ldots & HD\,145127 \\ 
  USco\,1610$-$1913B & 16:10:32.33 & -19:13:08.67      & 2011 June 19 & 720 & M9.0$\pm$0.5 & 5.8 & 115.4$^{\circ}$   & 12.74$\pm$0.01 & HD\,145127 \\ 
  \ldots & \ldots & \ldots & 2013 April 16 & 360 & \ldots & \ldots & \ldots & \ldots & HD\,144925 \\ 
  USco\,1610$-$2502B & 16:10:18.87 & -25:02:32.78      & 2011 June 19 & 960 & M5.5$\pm$0.5 &  5.1 & 239.3$^{\circ}$  & 11.25$\pm$0.01 & HD\,145127 \\ 
  HIP\,77900B       & 15:54:30.47 & -27:19:57.51       & 2011 June 22 & 720 & M9.0$\pm$0.5 & 21.8 & 12.7$^{\circ}$   & 14.04$\pm$0.01 & HD\,146606 \\
  USco\,1612$-$1800B & 16:12:48.97 & -18:00:49.56      & 2012 July 7  & 300 & M8.5$\pm$0.5 & 3.0 & 11.0$^{\circ}$    & 13.20$\pm$0.01 & HD\,144925 \\ 
  & & & &                           & & & & & \\
  \multicolumn{10}{c}{Background Objects} \\
  & & & &                          & & & & & \\
  HIP\,78099-2B                  & 15:56:48.53 & -23:11:10.69 & 2011 June 20 & 960 & reddened early-type star   & 12.4 & 132.0$^{\circ}$  & 15.72$\pm$0.02 & HD\,145127 \\
  2MASS\,J16141484$-$24270844-6B & 16:14:14.71 & -24:27:06.23 & 2011 June 20 & 960 & reddened M~star            & 2.8 & 320.6$^{\circ}$   & 15.24$\pm$0.02 & HD\,145127 \\
  USco\,160936.5-184800-7B       & 16:09:36.51 & -18:47:55.52 & 2011 June 21 & 600 & reddened early-type star   & 5.5 & 356.7$^{\circ}$   & 14.67$\pm$0.01 & HD\,138813 \\
  GSC06794-00537-18B             & 15:50:58.13 & -25:45:28.05 & 2012 July 5 & 480 & reddened early-type star    & 52.5 & 26.1$^{\circ}$   & 17.95$\pm$0.19 & HD\,145188 \\
  USco\,16213591$-$23550341-2B   & 16:21:35.37 & -23:54:38.43 & 2012 July 5 & 480 & reddened early-type star    & 26.3 & 343.5$^{\circ}$  & 14.66$\pm$0.01 & HD\,144254 \\
  USco\,161437.5$-$185824-0B     & 16:14:37.15 & -18:58:43.13 & 2012 July 7 & 480 & reddened early-type star    & 19.7 & 195.3$^{\circ}$  & 16.14$\pm$0.03 & HD\,144925 \\
  USco\,161437.5$-$185824-4B     & 16:14:35.87 & -18:58:19.09 & 2012 July 7 & 480 & reddened early-type star    & 24.0 & 282.1$^{\circ}$  & 15.94$\pm$0.03 & HD\,144925 \\
  DENIS\,P\,J162041.5$-$242549-5B & 16:20:42.34 & -24:25:35.59 & 2012 July 8 & 480 & reddened early-type star   & 18.3 & 41.9$^{\circ}$   & 15.51$\pm$0.02 & HD\,142705 \\
  USco\,160606.29$-$233513.3-1B  & 16:06:07.48 & -23:34:57.16 & 2012 July 8 & 480 & galaxy                      & 22.9 & 45.3$^{\circ}$   & 15.77$\pm$0.03 & HD\,142705 \\
  & & & &                          & & & & & \\
  \enddata
  \tablenotetext{a}{2MASS magnitudes used because UKIDSS photometry was unavailable.}
  \tablenotetext{b}{The ellipses (\ldots) signify that the value is the same as in the row above.}
\end{deluxetable}

\clearpage
\end{landscape}



\clearpage

\begin{landscape}
\begin{deluxetable}{lccccc}
  \tablecaption{Upper Sco New Companions \label{table:tab2}}
  \tabletypesize{\tiny}
  \tablewidth{0pt}
  \tablehead{
    \colhead{Property} &
    \colhead{HIP\,77900B} &
    \colhead{USco\,1610$-$1913B\tablenotemark{a}} &
    \colhead{USco\,1612$-$1800B} &
    \colhead{USco\,1602$-$2401B\tablenotemark{a}} & 
    \colhead{USco\,1610$-$2502B\tablenotemark{a}} 
  }
  \startdata
                 \multicolumn{6}{c}{{\bf Directly Measured Properties}} \\
          \hline \\
                  RA (J2000) &          15:54:30.47 &         16:10:32.33  &          16:12:48.97  &          16:02:51.17  & 16:10:18.87 \\
                  Dec (J2000) &         -27:19:57.51 &        -19:13:08.67 &         -18:00:49.56 &          -24:01:50.45   & -25:02:32.78 \\
          Separation (AU) &       3200\,$\pm$\,300 &          840\,$\pm$\,90 &       430\,$\pm$\,40 &     1000\,$\pm$\,140  & 730\,$\pm$\,80 \\
          Primary Name & HIP\,77900 & USco\,161031.9$-$191305 & USco\,161248.9$-$180052 & USco\,1602.8$-$2401 & USco\,161019.18$-$250230.1 \\
          Primary SpT\tablenotemark{b} & B6$\pm$1 & K7 & M3 & K4 & M1 \\ 
                 SpT &        M9\,$\pm$\,0.5 &             M9\,$\pm$\,0.5 &                                M8.5\,$\pm$\,0.5 &                           M7.5\,$\pm$\,0.5 &                 M5.5\,$\pm$\,0.5 \\
                \ips\tablenotemark{c}  &             \ldots       &             18.31$\pm$0.10 &                                 18.48$\pm$0.06 &                           15.77$\pm$0.01  &                   15.12$\pm$0.19 \\
                \zps\tablenotemark{c}  &             \ldots       &                \ldots         &                                16.83$\pm$0.01 &                             14.8$\pm$0.2 &                   14.3$\pm$0.5 \\
                \yps\tablenotemark{c}  &             \ldots       &            15.81$\pm$0.02 &                                 16.10$\pm$0.04 &                              14.1$\pm$1.7  &                    13.6$\pm$1.1 \\
                 $Z$\tablenotemark{c}  &       16.86$\pm$0.01 &                \ldots         &                                      \ldots        &                                  \ldots       &                         \ldots      \\
                 $Y$\tablenotemark{c}  &      15.86$\pm$0.01 &                \ldots         &                                      \ldots        &                                  \ldots       &                         \ldots      \\
                 $J$\tablenotemark{c}  &      15.07$\pm$0.01 &           13.90$\pm$0.09\tablenotemark{d}  &                    \ldots        &            12.50$\pm$0.06\tablenotemark{d} &  12.18$\pm$ 0.06\tablenotemark{d} \\
                 $H$\tablenotemark{c}  &      14.52$\pm$0.01 &           13.30$\pm$0.01 &                                13.73$\pm$0.01 &                            11.97$\pm$0.01 &                11.65$\pm$0.01 \\
                 $K$\tablenotemark{c}  &      14.04$\pm$0.01 &           12.74$\pm$0.01 &                                13.20$\pm$0.01 &             11.60$\pm$0.05\tablenotemark{d} &               11.25$\pm$0.01 \\
               $J-H$ &      0.54$\pm$0.02 &                \ldots         &                                       \ldots       &              0.53$\pm$0.06\tablenotemark{d} &               0.60$\pm$0.01  \\
               $H-K$ &      0.49$\pm$0.02 &           0.56$\pm$0.02 &                                0.53$\pm$0.04 &              0.37$\pm$0.05\tablenotemark{d} &               0.40$\pm$0.01  \\
         $K_{S}$-$W4$ &            \ldots        &                 \ldots        &                                    \ldots          &               $\le$\,3.06\tablenotemark{d}  &               4.9$\pm$0.2      \\
 Primary $K_{S}$-$W4$ &        -0.26$\pm$0.09 &             1.1$\pm$0.2  &  2.2$\pm$0.3 ($\lesssim$\,1.8)\tablenotemark{e} &          $\le$\,0.98\tablenotemark{d}   &               $\le$\,1.2     \\
                & & & & & \\
          \hline \\
                                  \multicolumn{6}{c}{{\bf Derived Properties}} \\
                 & & & & & \\
          \hline \\
          Primary $M_{bol}$ &                  -0.72$^{+0.20}_{-0.20}$\tablenotemark{f} &           2.17$^{+0.07}_{-0.07}$\tablenotemark{f} &         2.40$^{+0.13}_{-0.13}$\tablenotemark{g} &                 2.57$^{+0.02}_{-0.02}$\tablenotemark{f} &                    0.93$^{+0.12}_{-0.12}$\tablenotemark{g} \\
          Primary $T_{eff}$ &                  13700$^{+1550}_{-1200}$\tablenotemark{f} &             4140$^{+130}_{-160}$\tablenotemark{f} &         3410$^{+130.}_{-150.}$\tablenotemark{g} &                     4550$^{+120}_{-120}$\tablenotemark{f} &                      3700$^{+150}_{-140}$\tablenotemark{g} \\
          Primary Mass (5Myr) &                3.8$^{+0.7}_{-0.5}$\tablenotemark{f} &           0.88$^{+0.14}_{-0.17}$\tablenotemark{f} &         0.36$^{+0.14}_{-0.12}$\tablenotemark{g} &                  1.34$^{+0.12}_{-0.13}$ \tablenotemark{f} &                    0.70$^{+0.20}_{-0.20}$\tablenotemark{g} \\
          Primary Mass (10Myr) &               3.8$^{+0.8}_{-0.5}$\tablenotemark{f} &           0.87$^{+0.11}_{-0.18}$\tablenotemark{f} &         0.36$^{+0.14}_{-0.15}$\tablenotemark{g} &                  1.18$^{+0.06}_{-0.07}$ \tablenotemark{f} &                     0.70$^{+0.18}_{-0.17}$\tablenotemark{g}  \\
                 & & & & &  \\
          $M_{bol}$\tablenotemark{h} &                          11.38$\pm$0.13 &                                  10.09$\pm$0.13 &                                     10.52$\pm$0.13 &                                        8.87$\pm$0.13 &                         8.47$\pm$0.13 \\
          $T_{eff}$\tablenotemark{h} &      2400$^{+150}_{-150}$ (2390.$^{+130}_{-130}$) &       2400$^{+150}_{-150}$ (2400$^{+140}_{-140}$) &        2550$^{+150}_{-150}$ (2450$^{+140}_{-140}$) &               2790$^{+150}_{-150}$ (2550$^{+150}_{-140}$) &                            3050$^{+150}_{-140}$  \\
          Mass (5Myr)\tablenotemark{i}&                        19$^{+7}_{-4}$ (18$^{+3}_{-3}$) &                       19$^{+7}_{-4}$ (19$^{+4}_{-3}$) &                         23$^{+12}_{-6}$ (20$^{+5}_{-3}$) &                         41$^{+20}_{-13}$ (24$^{+8}_{-4}$) &                       100$^{+80}_{-50}$ \\
          Mass (10Myr)\tablenotemark{i} &                       20$^{+7}_{-3}$ (20$^{+4}_{-2}$) &                       20$^{+7}_{-3}$ (20$^{+4}_{-2}$) &                         26$^{+16}_{-7}$ (20$^{+6}_{-2}$) &                         47$^{+20}_{-18}$ (25$^{+10}_{-6}$) &                      100$^{+70}_{-40}$ \\
  \enddata
  \tablenotetext{a}{Proper motion confirmed companion from \citet{2009ApJ...703.1511K}.}
  \tablenotetext{b}{Spectral type for HIP\,77900 from \citet{1967ApJ...147.1003G}, for USco\,1610$-$1913 and Usco\,1612$-$1800 from \citet{2002AJ....124..404P}, USco\,1602.8$-$2401 from \citet{1999PhDT........89K}, and USco\,1610$-$2502 from \citet{1998AA...333..619P}.}
  \tablenotetext{c}{For all magnitudes we assume a minimum magnitude error of 0.01\,mag (see Section~\ref{sec:data}). The ellipses (\ldots) are used if no detection in that filter was available.}
  \tablenotetext{d}{2MASS magnitudes used because UKIDSS photometry was unavailable.}
  \tablenotetext{e}{$K_{S}$ and {\it WISE} photometry is of the primary due to unresolved separation, $\sim$\,3\arcsec. In this case, the $K_{S}$ is likely accurate since it agrees well with the UKIDSS photometry (which resolves the primary and companion). In parenthesis is the lower limit to the $K_{S}-W4$ color assuming the typical $K_{S}-W4$ color range for an M7.5 \citep{2012ApJ...758...31L} to extract its expected $W4$ magnitude.}
  \tablenotetext{f}{The uncertainties for HIP\,77900 result from an assumed spectral type uncertainty of $\pm$\,1 subclass. Mass uncertainties in USco\,161031.9-191305, USco\,1602.8-2401, and USco\,161248.9-180052 include 124\,K temperature uncertainty from the spectral type conversion and an assumed spectral type uncertainty of half a subclass. We derive $M_{bol}$ using bolometric corrections relationship to spectral type from \citet{1982_Schmidt-Kaler}, $T_{eff}$, and mass derived using the \citet{2000AA...358..593S} evolutionary models. See Section~\ref{sec:results} for details.}
  \tablenotetext{g}{Mass derived with the $T_{eff}$ and the \citet{1998AA...337..403B} evolutionary models. The mass derived using the \citet{2003ApJ...593.1093L} temperature scale. The errors in the mass come from the uncertainty in temperature, which includes uncertainties due to spectral type and the conversion between spectral type and $T_{eff}$}.
  \tablenotetext{h}{$M_{bol}$ (uncertainty of 0.13\,mag) is derived from empirical equations for the $K$ bolometric correction from \citet{2004AJ....127.3516G}. The $T_{eff}$ is derived using the temperature scale for young stars \citep{2003ApJ...593.1093L} where we assume an uncertainty of 124\,K in this scale \citep[same as in ][]{2004AJ....127.3516G}. We also show $T_{eff}$ from \citet{2004AJ....127.3516G}, which was derived for older field ultracool dwarfs (uncertainty of 124\,K), in parenthesis. The final uncertainty in $T_{eff}$ includes uncertainties due to spectral type and the conversion between spectral type and $T_{eff}$}
  \tablenotetext{i}{Mass derived with the $T_{eff}$ and models from \citet{2000ApJ...542..464C}. The mass derived using the \citet{2004AJ....127.3516G} temperature scale is in parenthesis. The errors in the mass come from the uncertainty in temperature and spectral type. For USco\,1612$-$1800B and USco\,1610$-$1913B the lower limit on the mass may be underestimated because it is extrapolated from the spectral type upper limit (M9.5) which is outside the \citet{2003ApJ...593.1093L} temperature scale (only M0$-$M9). Note for USco\,1610$-$2502B we use the \citet{1998AA...337..403B} evolutionary models for the mass. See Section~\ref{sec:results} for details.}

\end{deluxetable}   

\clearpage
\end{landscape}




\clearpage

\begin{landscape}
\begin{deluxetable}{lcccccccc}
  \tablecaption{Average SEDs for Cluster Dwarf Stars \label{table:tab3}}
  \tabletypesize{\small}
  \tablewidth{0pt}
  \tablehead{
    \colhead{Spectral} &
    \colhead{\ips} &
    \colhead{\zps} &
    \colhead{\yps} &
    \colhead{$Z$} &
    \colhead{$Y$} &
    \colhead{$J$} &
    \colhead{$H$} &
    \colhead{$K$}
\\
    \colhead{Type} &
    \colhead{(AB)} &
    \colhead{(AB)} &
    \colhead{(AB)} &
    \colhead{(Vega)} &
    \colhead{(Vega)} &
    \colhead{(Vega)} &
    \colhead{(Vega)} &
    \colhead{(Vega)} \\
  }
  \startdata
               M0 &                                   \ldots &                  7.70$\pm$0.16 (10) &                   7.4$\pm$0.3 (12) &                   6.8$\pm$0.3 (12) &                   6.5$\pm$0.3 (12) &                   6.12$\pm$0.19 (11) &                   5.60$\pm$0.08 (3) &                   5.30$\pm$0.19 (11) \\ 
             M0.5 &                   8.36$\pm$0.06 (3) &                    7.75$\pm$0.18 (13) &                   7.5$\pm$0.2 (16) &                   6.99$\pm$0.19 (15) &                   6.72$\pm$0.17 (13) &                   6.3$\pm$0.18 (15) &                    5.81$\pm$0.12 (3) &                   5.43$\pm$0.17 (17) \\ 
               M1 &                   8.58$\pm$0.15 (19) &                   8.1$\pm$0.2 (26) &                   7.8$\pm$0.3 (31) &                   7.2$\pm$0.3 (32) &                   7.0$\pm$0.3 (29) &                   6.5$\pm$0.2 (25) &                   5.99$\pm$0.11 (13) &                   5.6$\pm$0.3 (32) \\ 
             M1.5 &                   8.9$\pm$0.2 (27) &                   8.4$\pm$0.3 (36) &                   8.0$\pm$0.4 (36) &                   7.5$\pm$0.4 (38) &                   7.2$\pm$0.3 (32) &                   6.8$\pm$0.3 (33) &                   6.2$\pm$0.2 (23) &                   5.9$\pm$0.4 (39) \\ 
               M2 &                   9.2$\pm$0.4 (19) &                   8.6$\pm$0.3 (23) &                   8.2$\pm$0.5 (23) &                   7.9$\pm$0.5 (20) &                   7.5$\pm$0.4 (19) &                   6.9$\pm$0.4 (20) &                   6.4$\pm$0.3 (16) &                   6.1$\pm$0.4 (24) \\ 
             M2.5 &                   9.5$\pm$0.5 (34) &                   8.9$\pm$0.5 (40) &                   8.5$\pm$0.5 (36) &                   8.2$\pm$0.4 (34) &                   7.8$\pm$0.4 (30) &                   7.3$\pm$0.4 (29) &                   6.7$\pm$0.4 (31) &                   6.4$\pm$0.4 (40) \\ 
               M3 &                   9.9$\pm$0.4 (45) &                   9.3$\pm$0.4 (59) &                   8.9$\pm$0.4 (55) &                   8.5$\pm$0.4 (64) &                   8.0$\pm$0.4 (59) &                   7.5$\pm$0.3 (58) &                   6.9$\pm$0.4 (62) &                   6.6$\pm$0.4 (66) \\ 
             M3.5 &                   10.4$\pm$0.5 (64) &                   9.8$\pm$0.4 (76) &                   9.3$\pm$0.4 (75) &                   9.0$\pm$0.4 (81) &                   8.5$\pm$0.4 (76) &                   7.9$\pm$0.4 (70) &                   7.4$\pm$0.4 (81) &                   7.1$\pm$0.4 (87) \\ 
               M4 &                   11.1$\pm$0.6 (56) &                   10.4$\pm$0.5 (78) &                   9.9$\pm$0.5 (75) &                   9.6$\pm$0.4 (88) &                   9.2$\pm$0.3 (85) &                   8.6$\pm$0.3 (80) &                   8.0$\pm$0.3 (89) &                   7.7$\pm$0.4 (93) \\ 
             M4.5 &                   11.7$\pm$0.4 (45) &                   10.9$\pm$0.4 (57) &                   10.4$\pm$0.3 (57) &                   10.1$\pm$0.3 (62) &                   9.6$\pm$0.3 (60) &                   9.0$\pm$0.3 (58) &                   8.4$\pm$0.3 (60) &                   8.1$\pm$0.3 (63) \\
               M5 &                 12.26$\pm$0.01 (2)  &                  11.45$\pm$0.07 (2)  &             10.89$\pm$0.01 (1)  &                                   \ldots &                                  \ldots  &                                \ldots         &         8.90$\pm$0.01 (1)  &             8.53$\pm$0.01 (1) \\
   \enddata
   \tablecomments{The number of objects used to calculate the
     average magnitude in each filter given in parenthesis after the
     average magnitude. SEDs constructed from only one object often
     have very small photometric errors (we assume a minimum of
     0.01\,mag, see Section~\ref{sec:data}) whereas the other SEDs will have large photometric
     errors because of the observed large spread in absolute magnitude
     in young stars.}
\end{deluxetable}

\clearpage
\end{landscape}



\clearpage

\begin{landscape}
\begin{deluxetable}{lcccccccc}
  \tablecaption{Average SEDs for Upper Sco Primaries \label{table:tab4}}
  \tabletypesize{\small}
  \tablewidth{0pt}
  \tablehead{
    \colhead{Spectral} &
    \colhead{\ips} &
    \colhead{\zps} &
    \colhead{\yps} &
    \colhead{$Z$} &
    \colhead{$Y$} &
    \colhead{$J$} &
    \colhead{$H$} &
    \colhead{$K$}
\\
    \colhead{Type} &
    \colhead{(AB)} &
    \colhead{(AB)} &
    \colhead{(AB)} &
    \colhead{(Vega)} &
    \colhead{(Vega)} &
    \colhead{(Vega)} &
    \colhead{(Vega)} &
    \colhead{(Vega)} \\
  }
  \startdata
               M0 &                    8.8$\pm$0.2 (2) &                   7.88$\pm$0.04 (1) &                    8.00$\pm$0.3 (2) &                                   \ldots &                                   \ldots &                                   \ldots &                      5.81$\pm$0.01 (1) &                                   \ldots \\ 
               M1 &                    8.7$\pm$0.1 (1) &                    8.5$\pm$0.7 (2) &                    8.3$\pm$0.8 (2) &                                   \ldots &                                   \ldots &                                   \ldots &                                   \ldots &                                   \ldots \\ 
               M2 &                    9.6$\pm$0.5 (4) &                    8.98$\pm$0.19 (3) &                    7.7$\pm$0.8 (8) &                                   \ldots &                                   \ldots &                                   \ldots &                    6.5$\pm$0.4 (4) &                     6.70$\pm$0.01 (1) \\ 
               M3 &                   9.3$\pm$0.6 (11) &                    8.4$\pm$0.7 (7) &                   7.6$\pm$0.7 (19) &                                   \ldots &                                   \ldots &                                   \ldots &                    6.4$\pm$0.6 (5) &                    6.3$\pm$0.7 (3) \\ 
             M3.5 &                    9.7$\pm$0.7 (2) &                    8.7$\pm$0.6 (2) &                    7.8$\pm$0.6 (5) &                     7.33$\pm$0.01 (1) &                     6.57$\pm$0.01 (1) &                    5.98$\pm$0.01 (1) &                   6.06$\pm$0.01 (2) &                    5.6$\pm$0.2 (3) \\ 
               M4 &                   9.6$\pm$0.7 (32) &                   8.6$\pm$0.7 (32) &                   7.9$\pm$0.8 (48) &                    7.5$\pm$0.3 (6) &                      7.0$\pm$0.3 (5) &                    6.3$\pm$0.3 (5) &                    6.5$\pm$0.7 (21) &                   6.1$\pm$0.9 (17) \\ 
             M4.5 &                   9.7$\pm$0.8 (14) &                   8.7$\pm$0.8 (14) &                   8.0$\pm$0.8 (24) &                                   \ldots &                     7.9$\pm$1.0 (3) &                    7.2$\pm$0.9 (3) &                     6.7$\pm$0.6 (9) &                    6.2$\pm$0.7 (8) \\ 
               M5 &                   9.8$\pm$0.8 (66) &                   8.8$\pm$0.8 (67) &                   8.3$\pm$0.7 (71) &                    8.6$\pm$1.0 (4) &                      7.8$\pm$0.9 (4) &                    7.0$\pm$0.8 (4) &                    6.4$\pm$0.5 (53) &                   6.3$\pm$0.6 (26) \\ 
             M5.5 &                   10.3$\pm$0.7 (46) &                   9.2$\pm$0.7 (44) &                   8.7$\pm$0.6 (43) &                                   \ldots &                    7.26$\pm$0.01 (1) &                    6.41$\pm$0.01 (1) &                    6.6$\pm$0.5 (38) &                   6.4$\pm$0.5 (24) \\ 
               M6 &                   11.1$\pm$0.7 (43) &                   9.9$\pm$0.6 (42) &                   9.3$\pm$0.6 (43) &                    9.1$\pm$0.4 (5) &                    8.1$\pm$0.4 (9) &                    7.2$\pm$0.5 (7) &                     7.0$\pm$0.6 (41) &                   6.7$\pm$0.5 (34) \\ 
             M6.5 &                   11.7$\pm$0.6 (13) &                   10.2$\pm$0.6 (13) &                   9.7$\pm$0.6 (13) &                                   \ldots &                                   \ldots &                                   \ldots &                   7.4$\pm$0.6 (14) &                   7.1$\pm$0.5 (12) \\ 
               M7 &                    11.7$\pm$1.2 (16) &                   10.1$\pm$0.8 (14) &                   9.7$\pm$0.9 (16) &                                   \ldots &                                   \ldots &                                   \ldots &                   7.6$\pm$0.8 (13) &                   7.3$\pm$0.7 (11) \\ 
             M7.5 &                    12.1$\pm$0.7 (7) &                    10.6$\pm$0.6 (6) &                    10.0$\pm$0.6 (7) &                                   \ldots &                                   \ldots &                                   \ldots &                    7.8$\pm$0.4 (6) &                    7.1$\pm$0.3 (4) \\ 
               M8 &                    12.5$\pm$0.9 (7) &                     10.7$\pm$1.0 (8) &                    9.9$\pm$1.0 (8) &                     11.05$\pm$0.01 (1) &                    10.04$\pm$0.01 (1) &                     9.17$\pm$0.01 (1) &                    7.9$\pm$0.6 (6) &                    7.4$\pm$0.6 (6) \\ 
             M8.5 &                    13.7$\pm$0.3 (3) &                  11.80$\pm$0.01 (2) &                    11.2$\pm$0.2 (3) &                                   \ldots &                                   \ldots &                                   \ldots &                    8.6$\pm$0.2 (3) &                    8.02$\pm$0.18 (3) \\ 
               M9 &                    13.6$\pm$0.3 (2) &                    12.2$\pm$0.5 (3) &                    11.4$\pm$0.4 (3) &                                   \ldots &                                   \ldots &                                   \ldots &                    8.8$\pm$0.4 (3) &                    8.3$\pm$0.4 (3) \\ 
               L0 &                   15.3$\pm$0.6 (4) &                   13.9$\pm$0.7 (6) &                   13.0$\pm$0.6 (6) &                       12.17$\pm$0.01 (1) &                    10.84$\pm$0.01 (1) &                      9.84$\pm$0.01 (1) &                   10.0$\pm$0.6 (6) &                   9.3$\pm$0.6 (6) \\
               L1 &                    14.8$\pm$0.5 (2) &                    13.1$\pm$0.4 (2) &                    12.6$\pm$0.8 (3) &                                   \ldots &                                   \ldots &                                   \ldots &                    9.9$\pm$0.7 (4) &                    9.2$\pm$0.6 (4) \\
               L2 &                                \ldots    &                   13.2$\pm$0.11 (1) &                      12.21$\pm$0.01 (1) &                                  \ldots &                                  \ldots &                                  \ldots &                       9.31$\pm$0.01 (1) &             8.63$\pm$0.01 (1) \\
  \enddata
  \tablecomments{The number of objects used to calculate the
    average magnitude in each filter given in parenthesis after the
    average magnitude. SEDs constructed from only one object often
    have very small photometric errors (we assume a minimum of
    0.01\,mag, see Section~\ref{sec:data}) whereas the other SEDs will have large photometric
    errors because of the observed large spread in absolute magnitude
    in young stars.}
\end{deluxetable}

\clearpage
\end{landscape}



\clearpage

\begin{landscape}
\begin{deluxetable}{lccccccccc}
  \tablecaption{SEDs for Field Dwarf Stars \label{table:tab5}}
  \tabletypesize{\small}
  \tablewidth{0pt}
  \tablehead{
    \colhead{Name} &
    \colhead{Spectral Type} &
    \colhead{\ips} &
    \colhead{\zps} &
    \colhead{\yps} &
    \colhead{$Z$} &
    \colhead{$Y$} &
    \colhead{$J$} &
    \colhead{$H$} &
    \colhead{$K$}
\\
    \colhead{} &
    \colhead{} &
    \colhead{(AB)} &
    \colhead{(AB)} &
    \colhead{(AB)} &
    \colhead{(Vega)} &
    \colhead{(Vega)} &
    \colhead{(Vega)} &
    \colhead{(Vega)} &
    \colhead{(Vega)}
  }
  \startdata
         2MASS J08533619-0329321        &   M9  &     15.61 $\pm$      0.01  &     14.13 $\pm$      0.01  &         \ldots                &      \ldots                  &                            &     11.18 $\pm$      0.05  &     10.48 $\pm$      0.05  &      9.91 $\pm$      0.05 \\
         2MASS J14284323+3310391        &   M9  &       \ldots                  &     14.89 $\pm$      0.03  &     13.92 $\pm$      0.01  &      \ldots                  &                            &     11.91 $\pm$      0.03  &     11.27 $\pm$      0.03  &     10.72 $\pm$      0.03 \\
         2MASS J15010818+2250020        &   M9  &     16.12 $\pm$      0.01  &     14.67 $\pm$      0.02  &     13.74 $\pm$      0.01  &      \ldots                  &                            &     11.76 $\pm$      0.05  &     11.18 $\pm$      0.05  &     10.69 $\pm$      0.05 \\
         2MASS J00242463-0158201        & M9.5  &     16.32 $\pm$      0.01  &     14.83 $\pm$      0.01  &     13.92 $\pm$      0.03  &      \ldots                  &                            &     11.73 $\pm$      0.03  &     11.10 $\pm$      0.03  &     10.53 $\pm$      0.03 \\
         2MASP J0345432+254023          &   L0  &     18.50 $\pm$      0.02  &       \ldots                  &     15.99 $\pm$      0.01  &      \ldots                  &     15.32 $\pm$      0.10  &     13.84 $\pm$      0.05  &     13.20 $\pm$      0.05  &     12.66 $\pm$      0.05 \\
         SDSSp J225529.09-003433.4      &   L0  &     20.01 $\pm$      0.07  &       \ldots                  &      \ldots                   &     17.00 $\pm$      0.05  &                            &     15.50 $\pm$      0.05  &     14.80 $\pm$      0.05  &     14.28 $\pm$      0.05 \\
  \enddata
\tablecomments{We assume a minimum of $\sigma=0.01$\,mag for all photometry (see Section~\ref{sec:data}).}
\end{deluxetable}

\clearpage
\end{landscape}

\clearpage

\end{document}